\documentclass{emulateapj}

\newcommand{\fmmm}[1]{\mbox{$#1$}}
\newcommand{\scnd}{\mbox{\fmmm{''}\hskip-0.3em .}}
\newcommand{\scnp}{\mbox{\fmmm{''}}}
\newcommand{\mcnd}{\mbox{\fmmm{'}\hskip-0.3em .}}
\newcommand{\cote}{C\^ot\'e}

\slugcomment{Accepted for publication by The Astronomical Journal - July 1,
2008.}

\shorttitle{The nucleus of the Sagittarius dwarf galaxy and M54}
\shortauthors{Bellazzini et al.}

\begin{document}

\title{The nucleus of the Sagittarius dSph galaxy and M54: \\
       a window on the process of galaxy nucleation.}

\author{M. Bellazzini}
\affil{INAF - Osservatorio Astronomico di Bologna, via Ranzani 1, 40127,
Bologna, Italy}
\email{michele.bellazzini@oabo.inaf.it}

\author{R.A. Ibata}
\affil{Observatoire Astronomique, Universit\'e de Strasbourg, CNRS, 11, rue de l'Universit\'e, F-67000 Strasbourg, France}
\email{ibata@astro.u-strasbg.fr}

\author{S.C. Chapman\altaffilmark{2}}
\affil{Institute of Astronomy, Madingley Road, Cambridge 
CB3 0HA}\altaffiltext{1}{University of Victoria, Victoria, BC, V8W 3P6, Canada}
\email{schapman@ast.cam.ac.uk}

\author{A.D. Mackey}
\affil{Institute for Astronomy, University of Edinburgh, Royal Observatory, Blackford Hill, Edinburgh EH9 3HJ}
\email{dmy@roe.ac.uk}

\author{L. Monaco}
\affil{European Southern Observatory, Alonso de Cordova 3107, Casilla 19001, Santiago, Chile}
\email{lmonaco@eso.org}

\author{M.J. Irwin}
\affil{Institute of Astronomy, Madingley Road, Cambridge CB3 0HA}
\email{mike@ast.cam.ac.uk}

\author{N.F. Martin}
\affil{Max-Planck-Institut f\"ur Astronomie, K\"onigstuhl 17, D-69117 Heidelberg, Germany}
\email{martin@mpia-hd.mpg.de}

\author{G.F. Lewis}
\affil{Institute of Astronomy, School of Physics, A29 University of Sydney, NSW 2006, Australia}
\email{gfl@physics.usyd.edu.au}

\and

\author{E. Dalessandro\altaffilmark{2}}
\affil{Dipartimento di Astronomia, Universit\`a di Bologna, via Ranzani 1, 40127,
Bologna, Italy}\altaffiltext{2}{ASI, Centro di Geodesia Spaziale, contrada Terlecchia, I-75100, Matera, 
Italy}
\email{emanuele.dalessandr2@unibo.it}

\begin{abstract}
We present the results of a thorough study of the nucleus of the Sgr dwarf
spheroidal galaxy (Sgr dSph) and of the bright globular cluster M54 (NGC~6715) 
that resides within the same nucleus (Sgr,N).  We have obtained accurate radial
velocities and metallicity estimates for 1152 candidate Red Giant Branch
stars of Sgr and M54 lying within $\sim 9\arcmin$ from the center of the galaxy,
from Keck/DEIMOS and VLT/FLAMES spectra of the infrared Calcium II triplet. 
Using both velocity and metallicity information we selected two samples of 425 
and 321 very-likely members of M54 and of Sgr,N, respectively.
The two considered systems display significantly different velocity dispersion
profiles: M54 has a steeply decreasing profile from $r=0\arcmin$, 
where $\sigma\simeq 14.2$ km/s, to $r\simeq 3\mcnd5$ where it reaches
$\sigma\simeq 5.3$ km/s, then it appears to rise 
again to $\sigma\simeq 10$ km/s at $r\sim 7\arcmin$. In contrast Sgr,N
has a uniformly flat profile at $\sigma\simeq 9.6$ km/s over the whole 
$0\arcmin \le r\le 9\arcmin$ range. Using data from the literature we show that the
velocity dispersion of Sgr remains constant at least out to $r\sim 100\arcmin$ 
and there is no sign of the transition between the outer
flat-luminosity-profile core and the inner nucleus in the velocity profile.
These results - together with a re-analysis of the Surface Brightness profile of
Sgr,N and a suite of dedicated N-body simulations - provide very strong
support for the hypothesis
that the nucleus of Sgr formed independently of M54, which probably plunged to
its present position, coincident with Sgr,N, because of significant decay of 
the original orbit due to dynamical friction.

\end{abstract}


\section{Introduction}

The occurrence of stellar nuclei at the photometric center of several dwarf
elliptical galaxies has been the subject of statistical investigations since the 
pioneering study by \citet{bts85} and \citet{bts87}, 
and nucleated dwarf ellipticals (dE,N) 
have become a generally recognized and well studied class of galaxies in their
own right \cite[see][and references therein]{GW,ferbin}. In the last few years the
results from systematic studies performed with the instrumentation on board of
the Hubble Space Telescope (HST), has revolutionized the field, providing 
strong evidence
supporting the possibility of an intimate connection between the process of
nucleation and the process of galaxy formation as a whole. In particular:

\begin{itemize}

\item It was generally accepted that stellar nuclei occurred in a significant
but minor fraction of dwarf elliptical galaxies \cite[$\sim$
25\%][]{bts85}; on the other hand, the high resolution HST analysis of a large 
and well selected sample of Virgo dEs by \citet[][hereafter C06]{cote6} has
revealed that the fraction of dE,Ns ($f_n$) can be as high as 66\% $\le f_n\le$
87\%. Strong support to a significant upward revision of $f_n$ comes also from
the recent ground-based study by \citet{grant}.
Hence, nucleation appears as the natural status of dEs instead of an
exceptional occurrence (see C06 for details and discussion).

\item An HST study of a large sample of Sa-Sd galaxies by \citet{bok} found that
a similarly high fraction of stellar nuclei is found also among these late type
spirals, $f_n\simeq$ 77\%. In general, the most recent studies agree in finding
a fraction of nucleation larger than 50\% in any kind of galaxy 
\cite[see, for example][and references therein]{carollo,balc,fercot,bok07}, 
with the only exception of those brighter than $M_B\simeq -20.3$ mag 
\cite[][hereafter FC06]{fercot}. Hence, nucleation seems to occur very
frequently in both dwarf and giant galaxies 
\citep[see][and references therein, for a deeper discussion of the 
dwarf/giant dichotomy]{GG}.

\item Stellar galactic nuclei are found to obey the same scaling relation that
links Supermassive Black Holes (SBH) with their host galaxies 
\cite[][C06, FC06]{wh}. It has been suggested that 
``{\em ...SBHs and stellar nuclei are nothing but complementary incarnation of
Central Massive Objects - they likely share a common formation mechanism and
follow a similar evolutionary path...}'' (FC06). In this view, the
accumulation of a compact overdensity of baryons at the very center would be a
common process during galaxy formation, while its further evolution into a SBH
or a stellar nucleus would depend on various ''local'' circumstances.

\item The similarity between stellar nuclei, the brightest globular clusters and
the recently discovered Ultra Compact Dwarf (UCD) galaxies \citep{ucd} clearly
suggests possible links among these classes of stellar systems
\citep{freeman,lotz1,lotz4,bmr94,hase,lucky,bok07}. 
In particular it has been suggested either
that nuclei can form by merging of pre-existing globular clusters 
\cite[][and references therein]{lotz1},
or that some objects currently classified as globular clusters may be in fact
galactic nuclei, the only relics of the tidal shredding of their host
galaxies that once orbited the Milky Way and/or M31 
\cite[see][and references therein]{macsyd,lucky,BS}, or even that most/all
globulars were nuclei of Galactic building blocks 
\cite[i.e. dwarf galaxies, see][BS, and references therein]{freeman,frebland,
bok07}.  

\end{itemize}

Within the scenario briefly outlined above, the crucial relevance of any
independent observational fact that may provide some insight on the process of
nucleation is clearly manifest. The mechanisms for the formation of stellar
nuclei that have been suggested until now belong to two broad classes:
(a) the orbital decay and merging at the center of the parent galaxies of
pre-existing stellar systems (star clusters), and (b) the accumulation of gas
(of various origins) in the very center of a galaxy and its subsequent 
transformation into stars \cite[see][for a detailed description of various
proposed flavors of these classes of models]{grant}. 

At present, all the
observational constraints we have in hand come from the study of the integrated
properties of samples of distant galaxies/nuclei. The nearest galaxies that are
known to host a stellar nucleus are M32, M33 and NGC205, in the Local Group.
Even in the most favorable case of NGC205, the
structure of the nucleus can be studied in some detail but it can be
resolved into individual stars only partially, even with HST imaging \citep{n205}.
In this context the case of the Sagittarius dwarf spheroidal galaxy
\cite[Sgr dSph,][]{iba94}
constitutes an absolute {\em unicum} \cite[][hereafter M05a]{lnuc}.

The Sgr dSph is a dwarf satellite ($L\sim 2-5 \times 10^7 L_{\sun}$) of the Milky 
Way \citep{iba97} whose main body is
at $D_{\sun}=26.3\pm 1.8$ kpc from us and $R_{GC}\simeq 18.7$ kpc from the 
center of the Galaxy \citep{ltip}. 
It is devoid of gas \citep{burton} and its stellar content is dominated by an
intermediate-age 
\cite[$\sim 6-8$ Gyr old, see][and references therein]{mic_con} metal-rich 
\cite[$\langle Z\rangle\sim$ \slantfrac{2}{5} $Z_{\sun}$,][and references 
therein]{luves} population, while $\simeq$
10\% of old and metal poor stars are also present \citep{lbhb}. The Sgr dSph is
currently being destroyed by the Galactic tidal field \citep{iba97}. 
The tidally
stripped stars form a huge and coherent tidal tail system that can be 
observed over the whole celestial sphere 
\cite[Sgr Stream, see][and references therein]{ibata01,maj}. 
There are four globular clusters within the main body of the galaxy 
\citep{iba94,DCA,paul} and probably several others that have been stripped from the
main body and are now associated with the Sgr Stream \citep{mic3,mic3b,carra}.

Particularly noticeable, however, is the very bright 
\cite[$M_V=-10.0$,][]{harris}
metal-poor ($[Fe/H]=-1.55$) \citep{brown} globular cluster M54 (NGC6715),
located exactly at the center of the light distribution of the Sgr galaxy.
For this reason, immediately after the discovery of the galaxy it was
suggested that M54 could be the actual nucleus of Sgr dSph \citep{bm95,sl95}.
Some doubt on this hypothesis was advanced by \citet{DCA} based on the
circumstance that the integrated color of M54 is much bluer than that of its
host galaxy, while the opposite was generally believed to occur in dE,Ns; 
however \citet{sl97} noticed that there were exceptions to this rule. In fact,
the latest studies suggest that the opposite may be true 
\cite[][see also C06]{lotz4}.
Since the mean metallicity of M54 and of the population dominating the Sgr
galaxy differ by one full dex, their Red Giant Branch (RGB) stars (and in some
cases also their Horizontal Branch stars, HB) can be easily discriminated 
from each other in a Color Magnitude Diagram (CMD); at the same magnitude the
RGB of Sgr is much redder than that of M54. Selecting genuine Sgr stars in this
way \citet{ls00} and \citet{maj} independently found an overdensity of Sgr
stars that appear concentric with M54 and have a similar spatial scale.
Using the same technique and a very large optical photometry database M05a
demonstrated that Sgr has actually a nucleus of metal rich stars (in the
following we will refer to this structure as the {\em nucleus} of 
the Sagittarius galaxy, {\em Sgr,N} for brevity, if not otherwise stated). 
This stellar
structure has the same center as M54, to within the uncertainties, but it
displays a different surface brightness profile, suggesting a different origin
from the cluster. The strong incompleteness in the innermost 
$\sim 10\arcsec$ prevented M05a
from obtaining accurate estimates of the structural parameters of the Sgr 
nucleus: they derived $-10.0 \le M_V\le -7.6$, $\mu_V(0) \la 18.5$ 
mag/arcsec$^2$, and, with a
tentative fit of a \citet{king} model $r_c\la 0.21 \arcmin$ and 
$r_t\sim 17\arcmin$ (where C=log($r_t/r_c$)$\sim$ 1.9; we will refine these 
parameters in Sect.~2, below).
They also found that the observed properties of the nucleus of the Sgr galaxy 
were fully compatible with those of known nuclei of dwarf ellipticals.
Moreover, the recent study by \citet{siegel} has shown 
that while the stellar budget of Sgr,N is dominated by the same 
intermediate-age population found in the surrounding galaxy, there is
clear evidence of more recent and repeated episodes of star formation 
\cite[not observed in extra-nuclear fields,][]{mic_con}, pointing to a history of
subsequent phases of re-accumulation of gas followed by star formation bursts,
very similar to what is observed by \citet{rossa} in a large sample of 
distant galactic nuclei.

In any case, the key result of M05a is that Sgr is a nucleated galaxy
independently of the presence of M54; even if one were able to remove the cluster from the galaxy
by magic, a nucleus made of typical Sgr stars would still be there. To explain
the strict spatial coincidence of M54 and the nucleus, M05a proposed two
possibilities: (a) the nucleus formed {\em in situ} from processed Sgr gas
that has fallen to the bottom of the galactic potential well, and
M54 was (independently) driven into the same place by dynamical friction; (b)
M54 was born in the present place (or was driven there as in case a) and 
formed a co-spatial overdensity of Sgr stars by capturing them (or the gas
from which they formed) from the
surroundings. 
The key difference between the two hypothesis can be summarized by
the questions {\em ``did M54 provide the mass seed around which the metal-rich
nucleus of Sgr was assembled? is M54 the main contributor of the mass budget in
the innermost $\sim 100$ pc of the Sgr galaxy?''}, that enclose the basic 
point of hypothesis (b).

Hypothesis (a) was found to be fully compatible with all the
available information, while it was not possible to consider case 
(b) in more depth. In a recent paper, \citet{mibau} studied in detail the
process of the capture of ``field'' stars from an intervening star cluster and
they concluded that, even in the most favourable cases, the fraction of stars
captured in a Hubble time is negligible, less that $10^{-4}$ of the cluster
mass. The range of conditions studied by \citet{mibau} includes the ``M54 within
Sgr'' case, hence their conclusions are fully applicable here. Nevertheless,
the results by \citet{mibau} do not exclude the possibility that M54
operates as
a collector of processed gas at the center of Sgr.

However it seems quite reasonable to expect that if case (b) 
is true, the stars of the Sgr,N structure should share the same kinematics as 
M54 stars, or, at least, their kinematics should be compatible with a 
mass-follows-light model, or, finally, their kinematics should be different 
from that of extra-nuclear Sgr stars,
as they would be bound to M54 and would orbit within its potential \citep{gil}.
Note that M54 and Sgr,N are enclosed within the innermost $10\arcmin$ of a
galaxy whose core radius and limiting radius are as large as 
$r_c=224\arcmin\pm 10\arcmin$ and $r_t=1801\arcmin\pm 112\arcmin$, 
respectively \citep{maj}.
Hence, even if the main body of Sgr is clearly undergoing tidal disruption the
kinematics of the nuclear region should be unaffected and is consequently
expected to provide a trustworthy insight on the inner mass distribution
\citep{munoz}.

To follow-up these ideas and
to obtain a deeper insight into this unique case of a galactic nucleus that can
be studied {\em in vivo} on a star by star basis,
we performed an extensive spectroscopic survey,
principally aimed at the study of the kinematics of Sgr,N and M54. Here we
present the main results of this survey. 
The plan of the paper is as follows. As a preliminary step we re-analyse the
Surface Brightness profile of Sgr,N with new data in Sect.~2. In Sect.~3 we
describe the observation and data reduction of our spectroscopic survey, we test
the reliability and the accuracy of our measures, the observed 
metallicity distribution is briefly discussed and, finally, the selection of the
samples is described in detail. Sect.~4 is devoted to the kinematical analysis
of the M54 and Sgr,N samples and in Sect.~5 we briefly present the results of a
suite of N-body simulations aimed at the study of the orbital decay of M54
within the Sgr galaxy. Finally, in Sect.~6 we summarize and discuss 
our results.
\vskip0.9truecm

\section{Refining the structural parameters of Sgr,N}

To gain a better insight into the Surface Brightness (SB) profile of Sgr,N, we
reduced archive data obtained with the Wide Field Channel (WFC) of the 
Advanced Camera for Survey (ACS) on board of the Hubble Space Telescope (HST). 
This is a set of short and long exposure F814W and F606W images of the
center of M54, taken within the GO~10775 Treasury Program \citep{sara07}. 
The relative and absolute photometry of individual stars has been obtained 
following the same steps described in \citet{sollima} for data taken from the
same survey. Further details and the overall analysis of these data will be
presented elsewhere; an independent study of the stellar populations of M54 
and Sgr,N from the same images has been recently presented by \citet{siegel}. 
In the present context, we limit ourselves to complementing the SB profile of Sgr,N obtained
by M05a with a couple of points in the innermost $10\arcsec$, using the same
methodology as M05a, and, as a consequence, to obtain stronger constraints on
the structural parameters of Sgr,N.  

\begin{figure}
\plotone{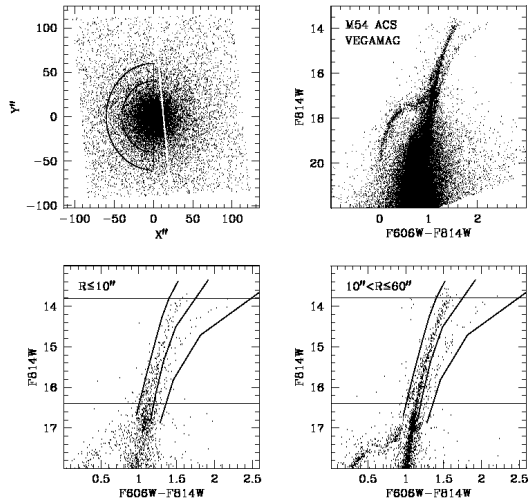}
\caption{ 
Upper left panel: the annuli and semi-annuli adopted for radial star counts are
plotted on the X,Y map of the stars having F814W$<20.0$; X,Y are the local
coordinates as defined in Sect. 3., below, but in arcseconds. 
Upper right panel:
overall CMD; note the difference among the steep and star rich RGB of M54 and
the redder and sparser RGB of Sgr,N, for  F814W$<17.0$.
Lower panels: the selection of M54 and Sgr,N RGB stars adopted as 
density tracers is illustrated for two radial ranges. 
The horizontal lines enclose
the magnitude range of the selection, the bluest and the reddest red lines
select the RGB stars of both systems and the middle red line separates the
sample into M54 members (bluer than the line) and Sgr,N members (redder than the
line).
\label{cmprof}}
\end{figure}

\begin{figure}
\plotone{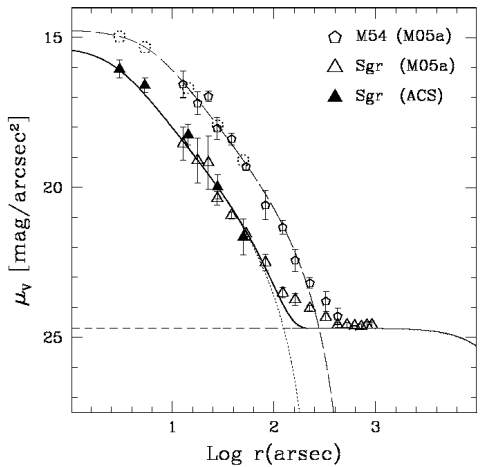}
\caption{Surface brightness profiles of M54 (pentagons) and of Sgr (triangles).
Continuous open symbols with errorbars are from M05a. The filled triangles have
been obtained from the density ratio computed here from HST/ACS data and are
normalized by adding them to the corresponding SB of the M54 best-fit model,
here marked by dotted pentagons.  The long dashed line is the best-fit
\citet{king} model found by \citet{tkd}  for M54; the short dashed line is the
best-fit King model found by \citet{maj} for the overall main body of the Sgr
galaxy; the dotted line is the adopted best-fit model for Sgr,N and the
continuous thick line is the sum of the last two models.
\label{prprof}}
\end{figure}

Fig.~\ref{cmprof} clearly shows that the new ACS photometry allows one to
discriminate the RGBs of M54 and Sgr,N by color even in the innermost
$10\arcsec$. Using the selection illustrated in Fig.~\ref{cmprof} we were able
to reliably measure the ratio of the density of tracer stars (RGB and AGB, in
the present case) in the two samples
in any given radial annulus $\rho_{Sgr,N}/\rho_{M54}$, as it scales as the ratio
of the number of stars per annulus  $N_{Sgr,N}/N_{M54}$. 
To avoid the inter-chip gap we used semi-annuli for $R>10\arcsec$.
The quantity 2.5log$(N_{Sgr,N}/N_{M54})$, in turn, scales as the difference in
Surface Brightness between the two systems 
\citep[see M05a for details, and][for the basis of the underlying evolutionary
flux theorem]{rbuz,rfp}\footnote{We used synthetic CMDs produced with the 
dedicated BASTI web interface \citep{cordier} to check that a population of age
= 12 Gyr and [Fe/H]=$-1.5$ (taken as a proxy for M54) and a population of age =
6 Gyr and [Fe/H]=$-0.4$ (taken as a proxy for Sgr,N) having {\em the same total
V luminosity}, places (approximately) the same number of stars in the selection
windows shown in Fig.~\ref{cmprof}.}. No attempt was made to estimate the
ellipticity of the system; as the system appears quite spherical at large radii
(see M05a) we assumed spherical symmetry. The coordinates of the centers of
M54 and Sgr,N, as derived from the $2\sigma$ clipped average X and Y of stars
selected by color as in Fig.~\ref{cmprof}, are found to coincide to within 
$< 2\arcsec$, in excellent agreement with M05a.
The effects of the radial variation of completeness on the
density ratio should be 
practically null, as we selected the M54 and Sgr,N samples in the same 
(bright) magnitude range ($13.8\le F814W\le 16.4$). 
To convert the estimated SB 
differences into an absolute scale we adopted the best-fit model profile of M54
by \citet{tkd} as a ``zero-point'', consistent with M05a. 
In practice, for each
observed $\Delta \mu^{Sgr-M54}(R_i)$ we obtained the surface brightness of Sgr,N
at radius $R_i$, $\mu^{Sgr}_V(R_i)$, as 
$\mu^{Sgr}_V(R_i)=\mu^{M54}_V(R_i)+\Delta \mu^{Sgr-M54}(R_i)$. The derived
portion of the profile is reported in Tab.~\ref{SSgr} and plotted in
Fig.~\ref{prprof}, together with the portions of the M54 and Sgr,N profiles
derived by M05a.

\begin{table}
\begin{center}
\caption{Surface Brightness profile of the 
         innermost $60\arcsec$ of Sgr \label{SSgr}}
\begin{tabular}{ccccc}
\tableline\tableline
$r_i$ & $r_f$ & r$_m$ & $\mu_V$ & e$_{\mu_V}$ \\
$\arcsec$ & $\arcsec$ & $\arcsec$ & mag/arcsec$^2$& mag/arcsec$^2$\\
\tableline
     0 &     5 &  3.01 & 16.1 & 0.3 \\  
     0 &    10 &  5.35 & 16.6 & 0.2 \\  
    10 &    20 & 14.30 & 18.3 & 0.3 \\  
    20 &    40 & 27.84 & 20.0 & 0.4 \\  
    40 &    60 & 49.78 & 21.7 & 0.6 \\  
\tableline
\end{tabular}
\tablecomments{$r_i$ and $r_f$ are the limits of the bins, $r_m$ is the average 
radius of the sample.}
\end{center}
\end{table}

\begin{table}
\begin{center}
\caption{Parameters of the King model that best fits the Sgr,N 
profile. \label{SSKing}}
\begin{tabular}{cccc}
\tableline\tableline
Parameter & Estimate & Uncertainty & Units\\
\tableline
 $r_c$      & 0.05  & $\pm$ 0.01 &arcmin\\  
 C          & 1.90  & \nodata & \nodata \\  
 $r_t$      & 4.00  & $\pm$ 0.8 & arcmin \\  
 $r_h$      & 0.42 & $\pm$ 0.08 & arcmin \\  
 $r_l$      & $\simeq$10.5 & \nodata & arcmin \\  
 $\mu_V(0)$ & 15.3  & $\pm$ 0.2 &mag/arcsec$^2$ \\  
 $M_V$      & -7.8  & $\pm$ 0.2 & mag\\  
\tableline
\end{tabular}
\tablecomments{$r_t$ is the tidal radius of the same King model, while $r_l$ is the observed
limiting radius, defined as the radius at which the observed profile of the
nucleus
appears to join the profile of the overall Sgr galaxy.
All the reported values and uncertainties have been estimated by assuming the
reported value of C.
The reported values of the surface brigthness are
not corrected for interstellar extinction.
$^a$ $r_h$ is the half-mass radius of the best fitting \citet{king} 
model: it can be considered equivalent to the half-light radius, 
in the present case. The half-light radius of the \citet{king62} best 
fitting the SB profile of Sgr,N is $r_h=0.23\arcmin$.}
\end{center}
\end{table}

The SB of Sgr,N is estimated at five radii. The three points at $R>10\arcsec$
nicely overlap with the observations by M05a, showing that the two profiles are
in good agreement. The two points at $R<10\arcsec$ provide very strong new
constraints on the fit of the overall profile. A C=1.90 King model with
$\mu_V(0)=15.4$ and $r_c=3.0\arcsec$ provides a good fit of the profile from the
innermost observed point, at $R\simeq 3\arcsec$, to 
$R\sim 120\arcsec = 2\arcmin$; 
for $R>2\arcmin$ the observed profile shows a much gentler decline with respect
to the model, joining the flat profile of the core of Sgr at $R=r_l\simeq
10.5\arcmin$ instead of $R=r_t\simeq 4.0\arcmin$. While there is no particular
physical reason in adopting a King model to represent Sgr,N, it provides an easy
and satisfying way to parametrize its inner SB profile. 
Moreover, while the best-fit
King model fails to reproduce the observed profile in the range 
$2\arcmin\la R\la 10.5\arcmin$, the fraction of the total Sgr,N light enclosed
in this range amounts to just a few per cent; therefore the adopted model
provides a reasonable description of the distribution of the bulk of the 
system light.   

The derived parameters are
listed in Tab.~\ref{SSKing}. Our results are fully compatible with the 
limits and the educated guesses by M05a; the most remarkable difference is in 
$r_c$ that was tentatively guessed as larger than that of M54 by M05a and, in
fact, is found to be significantly smaller here. The absolute integrated V
magnitude ($M_V=-7.8$) is just slightly brighter than the upper limit by M05
($M_V\le -7.6$). Assuming the best-fit models as correct, Sgr,N has a central
V luminosity surface density that is slightly more than half that of M54, 
its total V
luminosity is $\frac{1}{7}$ of the cluster and, finally both its core and 
half-light radii are $\sim$ one half of those of the 
cluster\footnote{$r_h$ is the half-mass radius of the best fitting \citet{king} 
model: it can be considered equivalent to the half-light radius, 
in the present case.  
Note that using the same definition, 
the half-light radius of M54 is $r_h=0.79\arcmin$, significantly larger than
the value reported by \citet[][$r_h=0.49\arcmin$]{harris}. The latter number is
very similar to what we obtain for the $r_h$ of the \citet{king62} empirical 
model that best-fits the profile of M54 (having the same C and $r_c$ adopted
here), i.e. $r_h=0.47\arcmin$. 
For C$\ge 1.0$ dynamical
\citet{king} models have $r_h$ larger than \citet{king62} empirical models by
more than 10\%; the difference becomes larger than a factor of two for 
C $\ge 2.0$. The half-light radius of the \citet{king62} best fitting the SB
profile of Sgr,N is $r_h=0.23\arcmin$. Hence independently of the adopted
family of models, the half-light radius of M54 is a factor of $\sim 2$ 
larger than Sgr,N one. Preliminary tests suggest
that available compilations of $r_h$ likely include quantities estimated in
various non-homogeneous ways.}.

With these new and much tighter constraints on the light distribution of Sgr,N
in hand, we will proceed in the following sections to perform a comparative study of its
kinematics and those of M54.

\section{The spectroscopic survey: observations, data reduction and selection of
the samples}

The spectroscopic observations on which this study is based were obtained with 
the multi-object
spectrographs DEIMOS on the Keck~2 telescope and FLAMES on the
VLT-UT2 telescope, as listed in Tab.~\ref{observations}. The DEIMOS observations were undertaken using two different
observing strategies: the first used the standard ``slitlet'' mode (the slitmask approach
used by the DEEP2 team, e.g.,  \citealt{davis03}), where 
short slits of minimum length $4\scnp$ and width $0\scnd7$ were milled; the
second mode used smaller  $1\scnd5$ long and $0\scnd7$ wide ``holes'' to allow
larger multiplexing. In both cases, the high resolution 1200~lines/mm
grating was employed with the OG550 blocking filter, giving a resolution of approximately $1.4\AA$ FWHM.
The slitlet spectra covered $\sim 6500$--$9000\AA$ and were extracted and wavelength
calibrated using the DEEP2 pipeline software; in contrast, the ``holes'' spectra
were extracted using software developed by our own group \citep{ibata05}, with the final extracted spectra
covering only the region $8400$--$8750\AA$ (purely due to a limitation in the software).

For the DEIMOS observations we selected candidate M54/Sgr,N RGB and Red Clump (RC) stars lying 
within $R\la 9\arcmin$ from the center of the systems from the wide-field
photometry of \citet{bump}. A limiting magnitude of ${\rm I=18}$ was adopted, 
which ensures that we probe the Red Giant Branch to approximately one magnitude below the Red Clump (see Fig.~\ref{seepri}).

Additional FLAMES observations were downloaded from the ESO archive; these targeted almost exclusively M54
stars, with only a small contamination from other populations. The FLAMES spectra were obtained with the high resolution 
setting HR21, which covers the calcium triplet from $8484$--$8757\AA$, with a resolution of $0.5\AA$ FWHM. 
The FLAMES data were extracted and calibrated using the ``girBLDRS'' pipeline\footnote{See {\tt http://girbldrs.sourceforge.net/}},
developed for the European Southern Observatory by Geneva Observatory.
Being of higher resolution, the FLAMES observations can be used to assess the accuracy of the DEIMOS
data, but more importantly FLAMES is a fiber-fed spectrograph and is therefore immune to the velocity errors that
can arise from mis-centering of slit spectrographs such as DEIMOS.

\begin{table}
\begin{center}
\caption{Spectroscopic observations. \label{observations}}
\begin{tabular}{cccc}
\tableline\tableline
Field & Instrument & Date & Exposure \\
\tableline
1 & FLAMES Medusa 1 & 28/07/2005 & $3 \times 845$~s \\
2 & FLAMES Medusa 1 & 28/07/2005 & $3 \times 845$~s \\
3 & FLAMES Medusa 2 & 28/07/2005 & $3 \times 845$~s \\
4 & FLAMES Medusa 1 & 20/08/2005 & $3 \times 795$~s \\
5 & DEIMOS holes & 31/08/2005 & $3 \times 900$~s \\
6 & DEIMOS holes & 31/08/2005 & $3 \times 900$~s \\
7 & DEIMOS holes & 02/10/2005 & $3 \times 950$~s \\
8 & DEIMOS holes & 02/10/2005 & $3 \times 950$~s \\
9 & DEIMOS holes & 03/10/2005 & $3 \times 720$~s \\
10 & DEIMOS holes & 27/05/2006 & $3 \times 300$~s \\
11 & DEIMOS holes & 27/05/2006 & $3 \times 300$~s \\
12 & DEIMOS slitlets & 28/05/2006 & $3 \times 720$~s \\
13 & DEIMOS slitlets & 23/09/2006 & $3 \times 300$~s \\
14 & DEIMOS slitlets & 25/09/2006 & $3 \times 300$~s \\
\tableline
\end{tabular}
\end{center}
\end{table}

The radial velocities of the target stars were obtained by cross correlating the
observed  spectra against an artificial template of the Ca~II lines, using
the same approach described and discussed in detail in \citet{ibata05} and
\citet{batta}. The determination of the 
uncertainty on the velocity is based on a weighted sum of Gaussian fit errors 
to the individual CaT lines as described in detail in the thorough discussion
by \citet{batta}, who also demonstrated that the velocity errors computed in
this way are reliable. An additional sanity check was provided by the comparison
with the scatter of the velocity measurements derived from the three 
individual Ca~II lines separately \cite[see, again,][]{batta}.

The metallicity of the target stars were estimated from the combined equivalent
width of the CaT lines. To this end we implemented the calibration described by
\citet{carrera},  which it is claimed works well up to high $[Fe/H]$ values.  
As
usual for CaT calibrations, [Fe/H] is function of the combined equivalent width
and of the difference between the V magnitude of the star and the mean level of
the HB/RR Lyrae of the system $V-V_{HB}$. For full consistency with  
\citet{carrera}, the equivalent widths of the CaT lines were measured according
to their prescriptions.
The metallicity measurements 
were put on the \citet{carretta} scale, 
assuming $V_{HB}=\langle V_{RR Ly} \rangle =18.17\pm 0.01$ and 
E(B-V)=0.14 \citep{ls00}.

\subsection{Photometric properties of the sample}

\begin{figure*}
\plotone{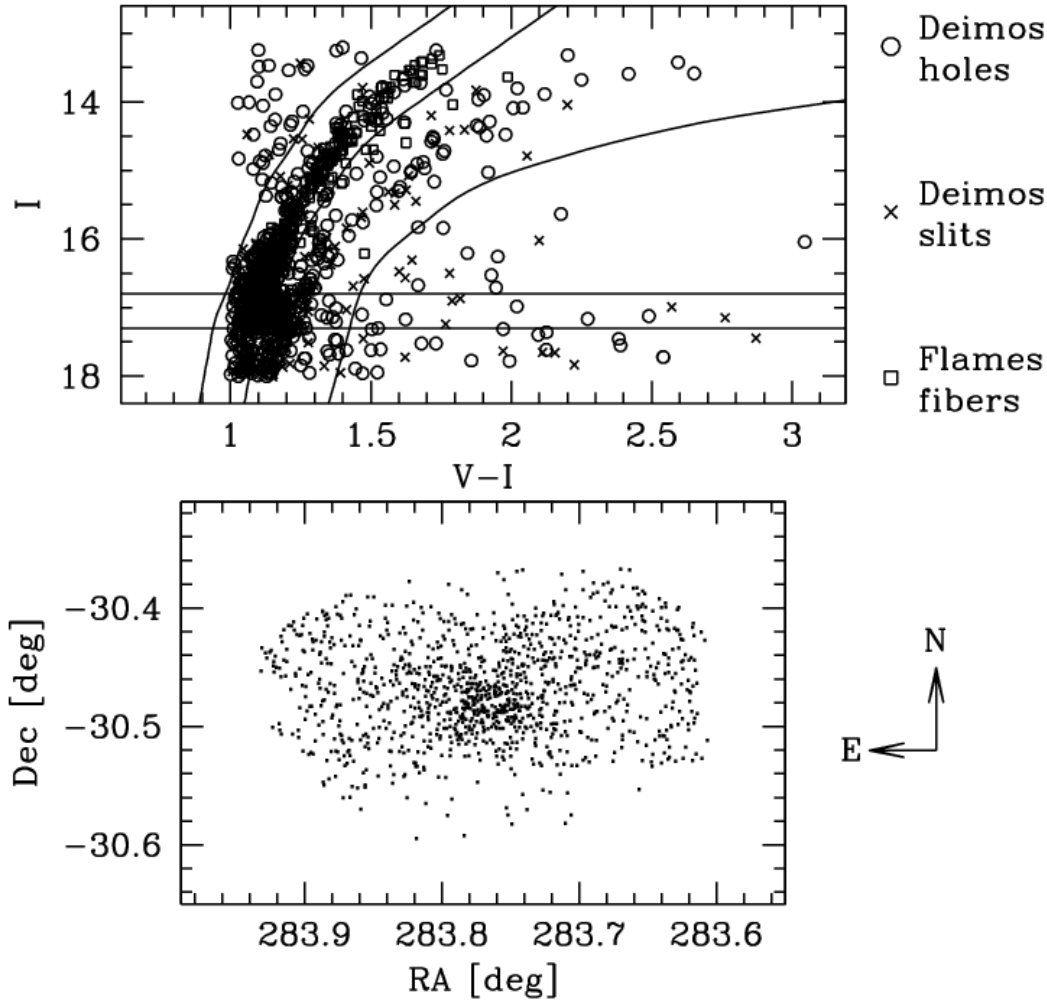}
\caption{CMD (upper panel) and distribution in the sky (lower panel) of the 1152
stars constituting our final sample. Different colors/symbols are associated to
measures obtained with different instruments and/or set-ups. The curves in the
CMD show the photometric selection box adopted for a first discrimination
between candidate M54 or Sgr,N members. 
\label{seepri}}
\end{figure*}

From these data we constructed a final catalogue containing the radial velocity and the
metallicity of 1152 stars in the innermost $15\arcmin \times 10\arcmin$ of the 
Sgr galaxy, as displayed in the lower panel of Fig.~\ref{seepri}. 
The CMD of all these stars is shown in the upper panel of the same figure. 
The RGB sequences of the two
systems are quite well separated in the CMD down to the faintest stars included
in our sample ($I\simeq 18$). 
However, we also selected stars lying on the Red Clump of
the Sgr,N population, at $V-I\simeq 1.15$ and $16.8< I< 17.3$: this feature
overlaps with the RGB of M54, hence we will need some further information to
establish the membership of these stars, in addition to their color, magnitude
and radial velocity. In particular, the metallicity would be very useful to
disentangle the two populations in this range. M54 has $[Fe/H]=-1.55$
\citep{brown,DCA}, with a small dispersion 
\cite[see][and Sect.~3.4, below]{sl95}. 
In contrast, the
metallicity distribution of the Sgr galaxy is dominated by a wide  
peak at $[Fe/H]\simeq -0.4$ 
\cite[see][and references therein]{luves,boni,mic_con},
extending from $[Fe/H]\sim -1.0$ to $[Fe/H]\sim 0.0$.
Stars more metal poor than $[Fe/H]\sim -1.2$ are quite rare and, presumably, 
they would pass their
core-Helium-burning phase as RR Lyrae or Blue Horizontal Branch stars
\citep{lbhb}: hence Red Clump stars of Sgr should have  $[Fe/H]\ga -1.0$, much
more metal-rich than any M54 star. 

As a first step for the selection of two samples representative of the Sgr,N and
M54 populations, we introduce the photometric 
classification defined by the
three curves overplotted on the CMD of Fig.~\ref{seepri}. Proceeding from left
(blue) to right (red) we assigned the flag {\em cmd} in the following way:
$cmd=1$ to the stars enclosed between the first and second curve, as likely M54
members, $cmd=2$ to the stars enclosed between the second and third curve, as 
likely Sgr,N members, and $cmd=0$ to stars lying to the blue of the first curve
or to the red of the third curve. A check a-posteriori has shown that all the
stars having $cmd=0$ have radial velocities incompatible with being members of
Sgr,N or M54, thus supporting the validity of our photometric classification.

As the coordinate of the centers of M54 and Sgr,N we adopt
$\alpha_0=283.763750\degr$ and $\delta_0=-30.478333\degr$ from \citet{ng6}, and we
convert to cartesian coordinates X,Y (in arcmin) projecting the equatorial
coordinates of each star ($\alpha$, $\delta$) on the plane of sky as 
in \citet{ven} 

\begin{displaymath}
X=-(10800/\pi)cos(\delta)sin(\alpha-\alpha_0)
\end{displaymath}
\begin{displaymath}
Y=(10800/\pi)[sin(\delta)cos(\delta_0)-cos(\delta)sin(\delta_0)cos(\alpha-\alpha_0)]
\end{displaymath}

\noindent
with X increasing toward the West and Y increasing toward the North.
Adopting the distance to Sgr and M54 measured by \citet{ltip}, 1~arcmin corresponds to
7.65~pc.

\subsection{Comparison between independent measures}

\begin{figure}
\plotone{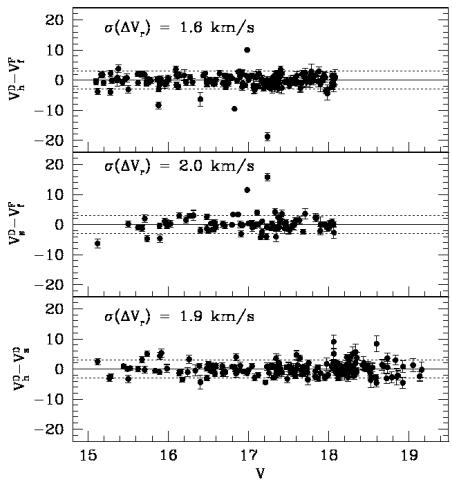}
\caption{Comparison between independent $V_r$ estimates obtained from spectra
acquired with different telescopes/instruments/set-ups. Each panel reports the
differences between the $V_r$ estimates as a function of V magnitude. The
$\pm 3.0$ km/s range around zero is enclosed by the two dotted lines. Upper
panel: DEIMOS/holes vs. FLAMES/fibers (160 stars); middle panel: 
DEIMOS/Slits vs. FLAMES/fibers (80 stars); lower panel: DEIMOS/holes vs. DEIMOS/Slits.
The standard deviation of each set of $\Delta V_r$, computed after recursive 
clipping of the very few 2.5-$\sigma$ outliers, is reported in each panel.
\label{compvr}}
\end{figure}

As there are several stars in common between the observational sets taken with
different instruments and or set-ups, we have the opportunity to check the
consistency and the accuracy of our $V_r$ and $[Fe/H]$ estimates.
In Fig.~\ref{compvr} we show the comparison between $V_r$ estimates from the
various sources: $V^{F}_f$ are the velocities obtained with FLAMES/fibers,  
$V^{D}_h$ are from DEIMOS/holes, and $V^{D}_s$ are from DEIMOS/slits.
It can be readily appreciated that the consistency among the different sets of
measures is excellent (i.e., $\Delta V_r\simeq 0.0$ km/s). 
The typical accuracy, as measured from the r.m.s. of the
$V_r$ differences is $\le \pm 2.0$ km/s, that is satisfying for the present
purpose. The actual uncertainty on the single measure should be a factor 
$\sqrt{2}$ smaller than the r.m.s. of $\Delta V_r$ (i.e. $\simeq \pm 1.4$ km/s), 
as the latter includes the uncertainties of both the estimates, 
added in quadrature.
Of the 414 velocity differences plotted in Fig.~\ref{compvr}, just
$\sim 10$ are significantly larger than $\pm 3.0$ km/s: some of these may be
associated with binary system observed at different orbital phases 
\cite[see][]{lstream}.

\begin{figure}
\plotone{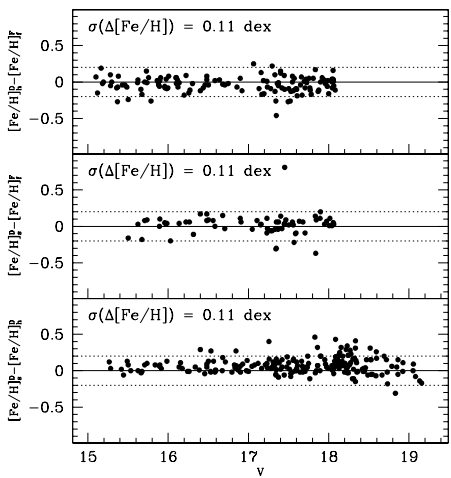}
\caption{Comparison between independent [Fe/H] estimates obtained from spectra
acquired with different telescopes/instruments/set-ups. Each panel reports the
differences between the [Fe/H] estimates as a function of V magnitude. The
$\pm 0.2$ dex range around zero is enclosed by the two dotted lines. Upper
panel: DEIMOS/holes vs. FLAMES/fibers; middle panel: 
DEIMOS/Slits vs. FLAMES/fibers; lower panel: DEIMOS/holes vs. DEIMOS/Slits.
The standard deviation of each set of $\Delta [Fe/H]$ is reported in each panel.
\label{compfe}}
\end{figure}

Fig.~\ref{compfe} shows the comparison between [Fe/H] estimates. In this case,
while the consistency between measures from FLAMES and DEIMOS/holes spectra is
very good, [Fe/H] from DEIMOS/slits spectra are $\sim 0.08$ dex larger - in
average - than those from the other two sources. We were unable to find the cause
of this mismatch and to preserve the homogeneity of the final merged dataset
we corrected all the [Fe/H] values from DEIMOS/slits spectra by adding $-0.08$
to all of them. The overall accuracy is more than satisfying, with a
typical r.m.s of 0.1 dex, and also in this case the outliers are rare.

\begin{figure}
\plotone{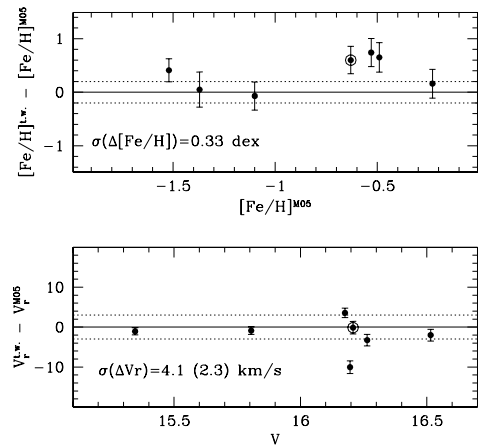}
\caption{Comparison between the [Fe/H] and $V_r$ estimates obtained in this work
(t.w.) and in \citet[][M05]{luves}, for the seven stars in common. The encircled
dot correspond to a star with molecular bands in its UVES spectrum, whose
metallicity is classified as tentative by \citet{luves}. In the lower panel, the
standard deviation value reported in parentheses refers to the sub-sample from
which the outlier at $\Delta V_r\simeq 10$ km/s has been removed.
The dotted lines encloses the range $0.0 \pm 0.2$ dex in the upper panel, and
$0.0 \pm 3.0$ km/s in the lower panel.
\label{compUVES}}
\end{figure}

Finally Fig.~\ref{compUVES} shows the comparison between our final $V_r$ and
[Fe/H] estimates and those obtained by \citet{luves} from high-resolution
FLAMES-UVES@VLT spectroscopy, for the seven stars in common between these datasets. The agreement in the
radial velocity is good: if the only outlier (at $\Delta V_r\sim -10 $ km/s) is
excluded the r.m.s. of the difference is just 2.3 km/s and the average
difference is zero\footnote{Also in this case we suggest multiplicity 
as the most likely origin
of the outlier \cite[see, again,][]{lstream}. Note that binaries are not
expected to significantly affect the observed velocity dispersions in systems
like those considered here \cite[see][for detailed discussion]{har,ols}.}. 
Therefore the reliability and the accuracy of our radial
velocity scale is fully confirmed. The comparison of metallicities is more
difficult to interpret and would have benefited by a larger number of stars in
common between the two samples. For four of the seven stars, ranging from
$[Fe/H]\simeq -1.5$ to $[Fe/H]\simeq -0.2$ the agreement is more than
acceptable, given the associated uncertainties. The other three stars show a
considerable $\Delta [Fe/H]$, but one of them was reported to have a 
tentative
[Fe/H] estimate from \citet{luves} because its spectrum was affected by
molecular bands making the analysis less reliable. In any case, none of
the observed differences are so large as to cause us to erroneously classify a 
Sgr,N
star as a M54 member, or vice-versa, according to the selection criteria that are
adopted below. Taking also into account the high degree of self-consistency
among different sets of measures shown in Fig.~\ref{compfe}, 
we conclude that also our metallicity scale is
sufficiently reliable and accurate for the purposes of the present analysis.

\subsection{Radial velocity distribution}

\begin{figure}
\plotone{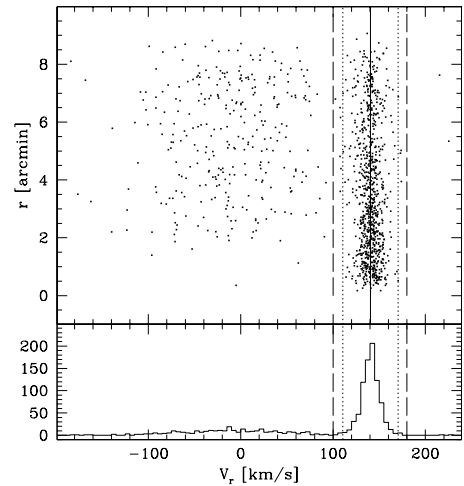}
\caption{Radial velocity of program stars as a function of distance from the
center of Sgr,N/M54 ({\em upper panel}), and radial velocity distribution 
({\em lower panel}). 
The long dashed lines mark the range we adopted for the first selection
of candidate Sgr/M54 members. The dotted lines enclose the 
(global) $\pm 3\sigma$ range
from the mean of the selected samples of candidates (continuous line). 
\label{vrsel}}
\end{figure}

The distribution of the radial velocity of all the observed stars as a function
of their (projected) distance ($r=\sqrt{X^2+Y^2}$) from the center of the
system is shown in Fig.~\ref{vrsel}. Two very different populations can be
identified in this plot: a broadly-distributed cloud of Galactic field 
stars showing a large dispersion around $V_r\sim 0$ km/s, and an  abundant
low-dispersion population with a systemic $V_r\sim 141$ km/s, typical of M54 and
Sgr,N. As a first broad selection, and following \citet{iba97}, 
we retained as possible M54/Sgr,N members 
all the (832) stars having 100 km/s $\le V_r\le$ 180 km/s 
(dashed lines in Fig.~\ref{vrsel}). 
We then computed the mean and the dispersion of this sample:
the $\pm 3\sigma$ range around the global mean is enclosed within the two
dotted lines. We do not reject the stars outside the $3\sigma$ range at this
stage, since the velocity distribution is different at different
radii (see below), and so a more refined choice is to exclude $>3\sigma$ outliers in any
considered radial bin. However, in Sect.~4 we will see that the adopted 
bin-by-bin $3\sigma$ rejection criterium will exclude all of these stars.

\begin{figure}
\plotone{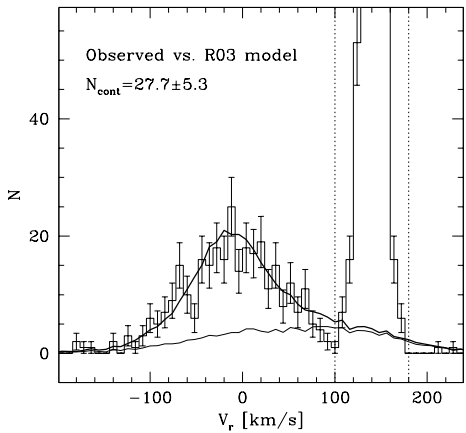}
\caption{Zoomed view of the observed radial velocity distribution (histogram
with  noise error bars)
around the peak of  the Galactic field population. The thick curve is the
distribution of all Galactic field stars with color and magnitude similar to
program stars predicted by the R03 Galactic model in the
considered direction, shifted by $-8$ km/s to match the position of the observed 
peak and rescaled to minimize the $\chi^2$ value with respect to
the observations for $V_r<100.0$ km/s. The thin curve shows the distribution of
(only) giant stars from the model, with the same normalization. 
The vertical dotted
lines enclose the 100 km/s $< V_r<$ 180 km/s interval. The expected number of
Galactic stars falling in this velocity range $N_{cont}$ is reported.
\label{besa}}
\end{figure}

Fig.~\ref{besa} shows that we expect some (limited) contamination 
from Milky Way stars even in our $V_r$ selected sample. 
The thick continuous line
superposed on the observed histogram of the radial velocities zoomed on the
Galactic component, shows the excellent agreement between the observations and
the predictions of the \citet{r03} Galactic model, once it has been 
shifted by $-8.0$ km/s to match the observed peak and rescaled to 
minimize $\chi^2$ for $V_r<100.0$ km/s.
The model suggests that up to $\sim 30$ Milky Way stars
may be present even in the relatively narrow $V_r$ window we have adopted to
select M54/Sgr,N stars. It is interesting to note that among the stars having
100 km/s $< V_r<$ 180 km/s but lying outside of the 
$\langle V_r\rangle \pm 3\sigma$ range, seven have 
$V_r<\langle V_r\rangle - 3\sigma$ and four have 
$V_r>\langle V_r\rangle + 3\sigma$, in agreement with the trend with $V_r$
predicted by the model.
Fig.~\ref{besa} shows also that most of the Galactic interlopers are expected to
be giants, according the the \citet{r03} Galactic model.
However, it is likely that such stars should be - in
general - much closer to us than Sgr: the model finds that the
contaminating giants have a mean distance of 8.7 kpc
(corresponding to $\Delta (M-m)=+2.4$~mag, with respect to Sgr), 
90\% of them have $D\le 11$
kpc and 99\% of them have $D\le 14$ kpc, i.e. much lower than $D_{Sgr}=26.3$
kpc. For this reason the difference between their V magnitude and $V_{HB}$ of
Sgr must be a bad proxy for their gravity.
The typical spectroscopic
metallicity error incurred by overestimating the distance modulus of these
Galactic stars by  2.4~mag is $\Delta {\rm [Fe/H]} = -0.58$, so the stars will
appear to be less metal-rich than they are in reality while
their color and magnitude mimic those of genuine Sgr/M54 stars. As a
consequence, for a Galactic star the metallicity obtained from the Calcium
Triplet (CaT) technique would normally be much different from what would be
expected from the magnitude and color measurements. Therefore, there is some
hope to exclude these Galactic interlopers from our final samples, identifying
them by their ``odd'' color-metallicity combination, as we will do in our final
selection (Sect.~3.5, below). As the mean metallicity of contaminating giants is
$\langle [Fe/H]\rangle = -0.75$ and 90\% of them have $[Fe/H]<-0.33$ it is quite
likely that many of them would have a measured metallicity lower than the
lower threshold we will assume for Sgr,N members ([Fe/H]$\ge -0.8$). On the other
hand those spuriously showing metallicities compatible with M54 will probably
be too red to be selected as possible cluster members (Sect.~3.5 and
Fig.~\ref{seepri}). 

\begin{figure*}
\plotone{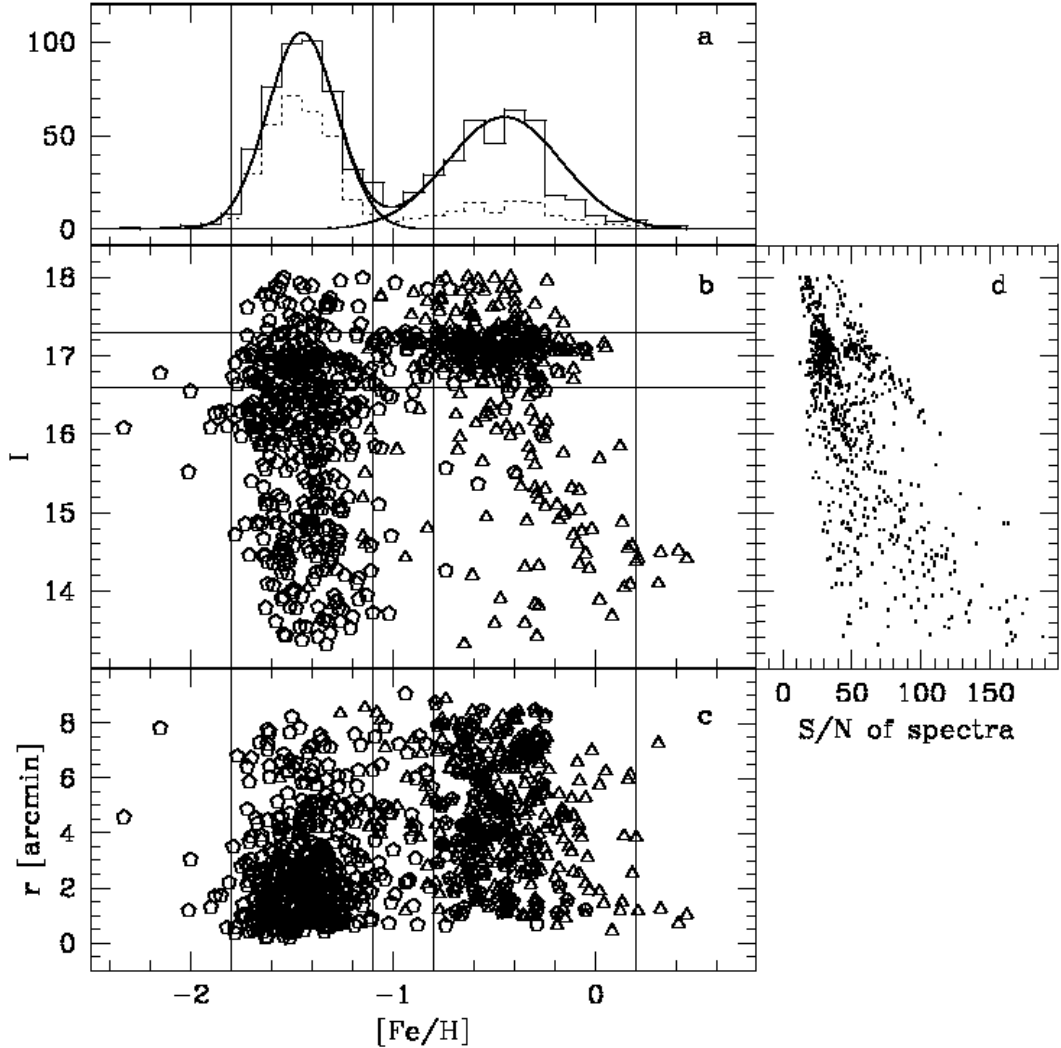}
\caption{Selection of clean samples of M54 and Sgr,N members.
{\em Panel a:} metallicity distribution of the 843 stars that survived 
the selection in radial velocity. The superposed curve is the sum of two
Gaussian distributions having mean and standard  deviation $(\langle
[Fe/H]\rangle, \sigma)=(-0.45,0.28)$ and $(-1.45,0.17)$, representing the
main populations of Sgr,N and M54, respectively. The dotted histogram is the
observed distribution for stars having $R\le 3.0\arcmin$.
{\em Panel b:} metallicity vs. I magnitude; open triangles are 
stars having $cmd=2$, i.e. photometrically selected as likely Sgr members (see
Fig.~\ref{seepri}); open pentagons are stars with $cmd=1$, i.e.
photometrically selected as likely M54 members.
The horizontal lines enclose the Sgr RC magnitude range ($16.6< I< 17.3$). 
The vertical lines mark the adopted thresholds in metallicity:
$-1.8\le [Fe/H]\le -1.1$, for the M54 sample;  $-0.8\le Fe/H]\le +0.2$ for 
the Sgr,N sample. 
{\em Panel c:} metallicity vs. distance from the center of M54.
The symbols are the same as in the above panel, except for asterisks superposed
on open pentagons: these are stars that lie on the RGB of M54 in the CMD but 
in fact belong to the superposed RC of Sgr, i.e. they have $16.6< I< 17.3$ and
a metallicity typical of  Sgr members. 
{\em Panel d:} Signal to Noise (per pixel) of the
spectra as a function of I magnitude. All the spectra have $S/N>12$.
\label{magsel}
}
\end{figure*}

Finally it is worth noting that
the spectra of all the 843 stars that passed the selection in $V_r$ have 
Signal to Noise ratio (per pixel) $S/N>12$; 300 of them have $S/N>50$ and 63 of
them have $S/N>100$ (see Fig.~\ref{magsel}d).

\begin{figure*}
\plotone{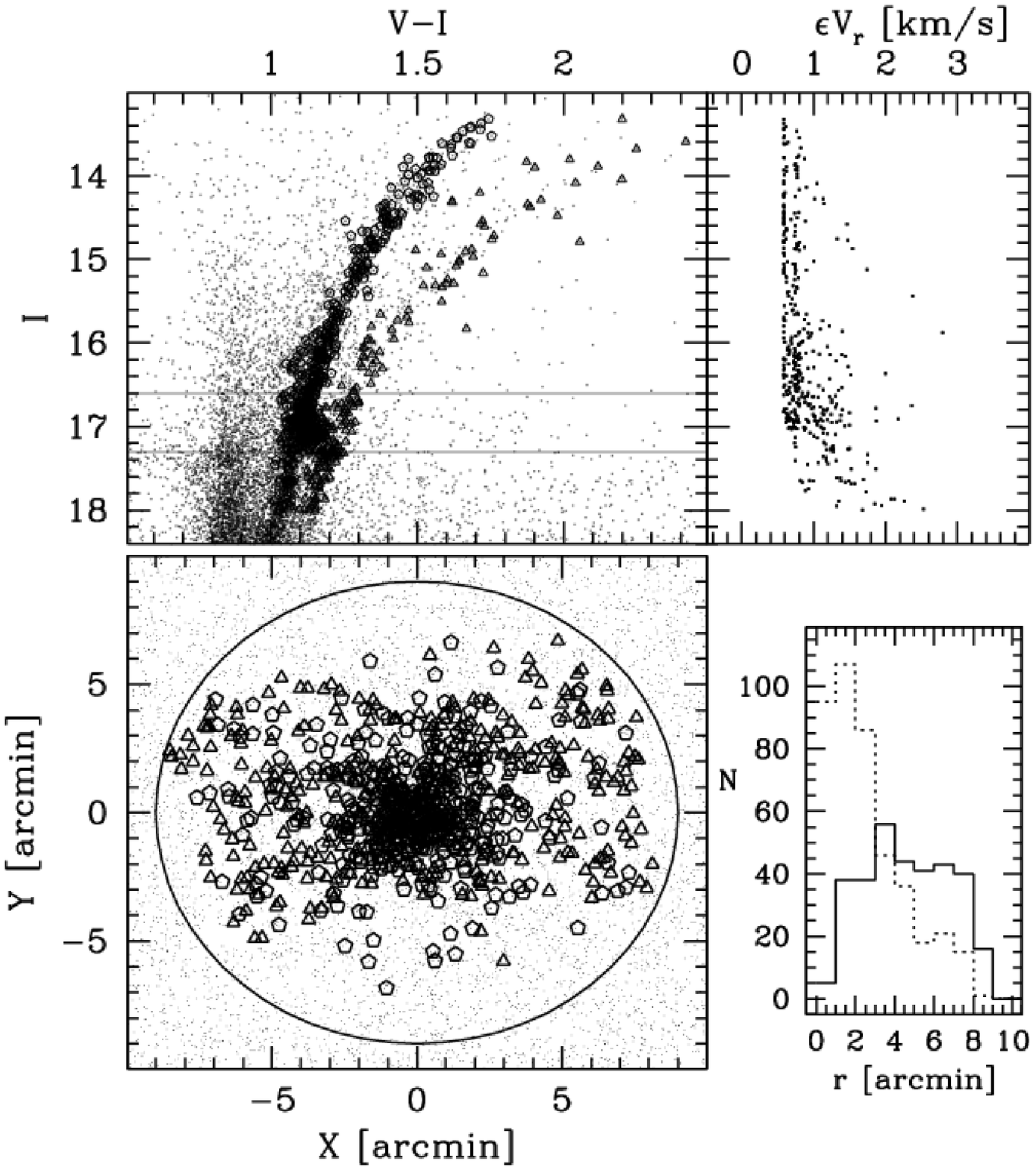}
\caption{{\em Upper left panel}: CMD of stars selected to belong to the
final Sgr,N (open triangles) and M54 (open pentagons) samples, superposed on the
CMD of all stars having $R<9\arcmin$ from \citet[][small dots]{bump} . 
The thin horizontal lines enclose the Red Clump of Sgr ($16.6< I< 17.3$).
The position in the sky of the selected stars relative to the center of 
Sgr,N/M54 is plotted in the {\em lower left panel}, the symbols are the same as
above. The radial distributions of stars selected on the Sgr,N sequences
(continuous histogram) and on the M54 sequences (dotted line) are plotted in
the {\em lower right panel}
The {\em upper right panel} shows the uncertainties of the $V_r$ estimates as a
function of I magnitudes for the selected stars. 
\label{cmsel}}
\end{figure*}

\subsection{Metallicity distribution}

The metallicity distribution of the velocity selected stars is shown if
Fig.~\ref{magsel}a. The presence of two populations is very clear: there is a
metal poor peak around $[Fe/H]\simeq -1.5$ that must be dominated by M54 stars
\citep{brown} and a broader distribution extending from $[Fe/H]\ga -1.0$ to
super-solar metallicity  corresponding to the main population of Sgr,N 
\cite[see][and references therein]{luves,mic_con,boni}. As mentioned earlier, in the present
context we will use the metallicity just as a further means to select samples of
M54 and Sgr,N members that are as clean as possible, so as to allow further analysis of their
kinematic properties. Therefore, we limit the discussion of the metallicity to the brief
considerations listed below.

\begin{enumerate}

\item{} We confirm that the bulk of Sgr stars have $[Fe/H]>-1.0$, belonging to a
broad distribution peaking around $[Fe/H]=-0.4$ and reaching super-solar
metallicities \citep{luves,boni}. \citet{lbhb} estimated that in the
main body of the Sgr galaxy old metal poor stars ($[Fe/H]\la -1.2$) should
provide less than 12\% of the whole stellar content. Several arguments suggest
that this fraction should be even lower in the very central region considered
here \cite[see][and references therein]{alard,mic_grad,siegel}.
However, we note that there are four stars having $[Fe/H]<-2.0$ in our sample
(see Fig.~\ref{magsel}) that are probably too metal poor to be members of M54
and might be part  of the metal-poor population of Sgr. 

\item{} The peak of the metallicity distribution corresponding to M54 is
best-fitted by a Gaussian curve having mean $\langle[Fe/H]\rangle =-1.45$. This
is slightly higher ($\sim +0.1$) than what was found by \citet{brown} and by
\citet{DCA}, while it is in good agreement with \citet{arma}. 
Note that any possible small shift in the zero point of our metallicity scale
does not affect the accuracy of the metallicity {\em ranking}, that is $\simeq
\pm 0.1$ and it is the relevant figure in the present context. 

\item{} Based on the analysis of the Color - Magnitude distribution of RGB stars,
\citet{sl95} concluded that there is an intrinsic metallicity spread among M54
stars of $\sigma_{int}([Fe/H])=0.16$ dex. Using spectroscopic metallicities
obtained with the CaT technique for five M54 members, \citet{DCA} found further
support for this hypothesis \cite[but see][]{brown}. 
Our much larger sample of CaT metallicities 
provides new supporting evidence for this possibility. The sample with
$-1.0\le [Fe/H]\le -2.0$ has an observed standard deviation of 0.17 dex,
larger than what is expected from the statistical scatter (see
Fig.~\ref{compfe}, above). Deconvolving the (internal) r.m.s. scatter 
(0.11 dex, see Fig.~\ref{compfe}) from the observed width of the distribution 
we
obtain an estimate of the intrinsic spread of $\sigma_{int}([Fe/H])=0.14$ dex,
in good agreement with previous estimates.  It seems very unlikely that the
width of the M54 peak is significantly contaminated by Sgr stars, as the width
remains unchanged if we limit the sample to the innermost $3\arcmin$ where
M54 should dominate the population mix. 
If finally confirmed by high resolution spectroscopy of a large sample of stars,
the purported metallicity spread would constitute another element of similarity
between M54 and other large-size very bright clusters like $\omega$ Cen, G1 and
the like, that are also suspected of being remnant nuclei of disrupted galaxies 
\cite[see][and references therein]{macsyd,lucky,marcio,sills,bok07}. 

\end{enumerate}

We think that the presently available metallicity
estimates provide a metallicity ranking that is more than sufficient for the
purposes of the present paper, i.e. (a) to discriminate M54 stars from Sgr,N 
stars in the magnitude range where the two population overlaps in the CMD 
($16.8< I< 17.3$), and (b) to exclude possible interlopers as stars having the
``wrong'' metallicity for their color.
The details of the adopted selections are described in the following 
sub-section. 

\subsection{The final selections}

The scheme of our finally adopted selections is illustrated in 
Fig.~\ref{magsel}b,c. Our aim is to have a
sample of M54 stars and a sample of Sgr,N stars as clean as possible from any
kind of interlopers. The adopted criteria are the following:

\begin{itemize}

\item{} We accept as members of the M54 sample the stars having 
$cmd=1$ {\bf and} $-1.8\le [Fe/H]< -1.1$. 

\item{} We accept as members of the Sgr,N sample the stars having 
$cmd=2$ {\bf and} $-0.8\le[Fe/H]\le +0.2$, or having $cmd=1$ and 
$-0.8\le[Fe/H]\le +0.2$ if $16.6<I<17.3$ (Sgr RC stars superposed on the RGB of
M54).

\end{itemize}

The metallicity separation avoids mixing between samples. The requirement
that the $cmd$ flag and the metallicity broadly agree in assigning the 
membership
is likely to exclude most of the Galactic interlopers. In the region in which
the RC of Sgr,N and the RGB of M54 overlap, the $cmd$ flag does not provide any
discrimination and we rely just on the metallicity. We have adopted quite
conservative criteria to obtain very reliable samples of genuine Sgr,N and M54
stars. As a further check, all the results presented in the following about the 
kinematics of Sgr,N have been verified to hold also when subsamples in
which stars in the RC region were excluded are adopted.
The main properties of the selected samples are shown in Fig.~\ref{cmsel}.

With the selection described above we select a sample of 425 
very likely M54 members and of 321 very likely Sgr,N members. 
In the process of clipping $3\sigma$ velocity outliers in any considered radial
bin (which will be performed in Sect.~4, below), we will further exclude eight 
stars from the M54 sample and three stars from the Sgr,N sample. Hence the actual
detailed analysis of the kinematics of the two systems will be performed on 417
stars for M54 and on 318 stars for Sgr,N.

\section{The kinematics of M54 and Sgr,N: observational facts}

In Tab.~\ref{m54dat} we list the estimates of some relevant physical parameters
of M54 that are available in the literature. For the structural
parameters ($r_c$, $r_t$, etc.) we will preferentially adopt those from
\citet{tkd}, since the best-fit model proposed by these authors is in good
agreement with the profile obtained from star counts in the outer
regions of the cluster by M05a. 
However in some
cases we will consider the effects of the adoption of different sets of
parameters on our results. It is reassuring that most authors derived
quite similar parameters from very different datasets. All the available mass
estimates for M54, also obtained with different methods and from different
datasets, range between 
$1.0\times 10^6 M_{\sun}$ and $2.0\times 10^6 M_{\sun}$
\citep{illi,mandu,tad,macgil,mcv}.

\begin{table*}
\begin{center}
\caption{Literature data for M54. \label{m54dat}}
\begin{tabular}{cccccccc}
\tableline\tableline
$r_c$& $r_t$  & C  & $\mu_V(0)$ & $V_t$ & [Fe/H]& method & Ref.\\
arcmin  & arcmin &    & mag/arcsec$^2$& mag & dex  \\
\tableline
0.11& 7.4& 1.83& 14.90& 7.61  &      &g.b. ap. phot. &\citet{illi2}\\
    &    &     &      & 7.68  &      &g.b. ap. phot. &\citet{peter}\\
0.11& 7.4& 1.84& 14.75&       &      &g.b. ap. phot. &\citet{tkd} \\
    &    &     &      &       &-1.55 &CaT spectr.    &\citet{DCA}\\
    &    &     &      &       &-1.55 &HR spectr.     &\citet{brown}\\
$\la$0.11&    &     &      &       &      &HST phot.      &\citet{macgil}\\
0.09& 9.9& 2.04& 14.35&       &      &g.b. ap. phot. &\citet{mcv}\\
0.05&    &     & 14.12&       &      &HST phot.      &\citet{ng6}  \\
\tableline
\end{tabular}
\tablecomments{g.b. ap. phot. = ground based aperture photometry; HST phot. =
integrated photometry and/or star counts from Hubble Space Telescope data;
CaT spectr. = medium resolution spectroscopy using the infrared Calcium Triplet
as metallicity indicator; HR spectr. = High Resolution spectroscopy and
elemental abundance analysis. See also \citet{web}, \citet{harris} and
\citet{macgil} for collections of literature data on M54.}
\end{center}
\end{table*}

In Tab.~\ref{Tvelmed} we report the average kinematical properties of the M54
and Sgr,N samples as a whole. The errors on all the reported quantities have been
obtained by a jackknife bootstrapping technique \citep{lupton}. 
The plain average $V_r$ of the two samples differs by $1.5\pm 0.7$ km/s, 
that is, just a $\sim 2\sigma$ difference. However an average with iterative
2-$\sigma$ clipping seems a much more appropriate estimator of the systemic
velocity since it is very robust to outliers and is more effective in estimating
the mode of the samples. The sigma clipping process converges very fast in
both cases, leaving very rich final samples ($N_{2\sigma}=246$ and $348$, for
Sgr,N and M54, respectively). The difference in the clipped mean velocities is 
a mere $0.8$ km/s. We conclude that the Sgr,N and M54 do coincide also in 
line of sight (l.o.s.) velocity, 
a further proof in support of their spatial coincidence. Many
authors have noted before that the systemic velocity of Sgr and M54 are very
similar \citep{iba97,DCA,luves}; we think we have {\em conclusively established}
here that M54 and {\em the nucleus of Sgr} have 
{\em the same} systemic l.o.s velocity, within $\simeq \pm 1.0$ km/s. 

\begin{table}
\begin{center}
\caption{Average kinematic properties of the M54 and Sgr,N samples. \label{Tvelmed}}
\begin{tabular}{lcccccc}
\tableline\tableline
   & N$_{tot}$ & $\langle V_r\rangle$ & $\sigma$      & $\sigma^{*}$  & N$_{2\sigma}$ & $\langle V_r\rangle_{2\sigma}$ \\
   &           &       km/s           &   km/s        &   km/s        &                    &              km/s                   \\
\tableline
Sgr,N& 321     & $139.4\pm 0.6$       & $ 10.0\pm 0.5$ & $ 9.6\pm 0.4$ &        246         &           139.9                     \\
M54  & 425     & $140.9\pm 0.4$       & $ 9.3\pm 0.5$  & $ 8.3\pm 0.3$ &        348         &           140.7                     \\
\tableline
\end{tabular}
\tablecomments{N$_{2\sigma}$ = number of stars in the sample once the 
iterative $2\sigma$
clipping algorithm to compute $\langle V_r\rangle_{2\sigma}$ has converged,
i.e. there is no more $2\sigma$ outlier. Note that at each iteration the algorithm
adopts the {\em current} value of $2\sigma$ that is, in general, significantly 
lower than the {\em global} value reported in the table.
$\sigma^{*}$ are the global values of the dispersion after the exclusion of the local 
$>3 \sigma$ outliers performed in Sect.~4.1. They are computed from the final cleaned samples
of 318 (Sgr,N) and 417 (M54) stars, as already reported in Sect.~3.5, above. }
\end{center}
\end{table}

The velocity dispersions of the whole Sgr,N and M54 do not seem very different,
at a first glance.
However, it should be recalled here that in the 
radial range covered by our data ($r\le 9\arcmin$), the surface brightness of 
M54 and Sgr,N declines by more than  $\sim 8$ mag/arcsec$^2$.
Hence, in the present context, what is really relevant
is the comparison between velocity dispersion profiles as a function of
distance from the center of the system, which will be presented in the 
following section.

\begin{figure}
\plotone{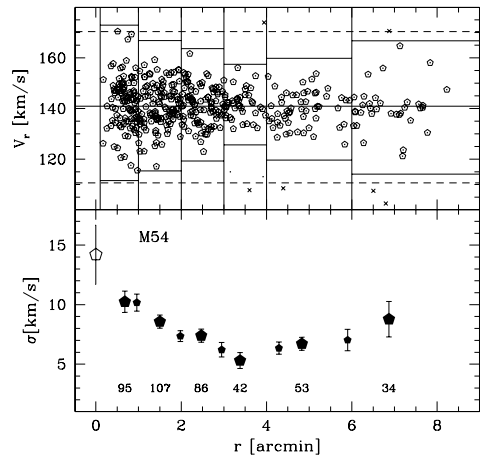}
\caption{Velocity dispersion profile of M54 stars. 
The {\em upper panel} shows the
$V_r$ distribution as a function of distance from the cluster center for
individual stars of the M54 sample. Only stars plotted as dots encircled by 
open pentagons are retained for the computation of $\sigma$ in the various
radial bins: small dots alone are stars rejected only because they are 
``local'' $3\sigma$ outliers of the bins, crosses are stars that would have 
been rejected also as $3\sigma$ outliers of the whole Sgr,N + M54 sample
(the global $\pm 3\sigma$ range is enclosed by the long-dashed lines).
The vertical lines display the adopted {\em independent} bins, of variable 
size. The {\em global} mean is marked 
by the continuous horizontal thick line. 
The {\em lower panel} displays the actual Velocity dispersion profile. 
The large filled pentagons are
the dispersions estimated in the corresponding bins displayed in the upper 
panel, with their bootstrapped errors. The number of stars per bin is also reported below the points.
The small filled pentagons are the estimates in the additional,
partially overlapping, bins.
The open pentagon is the estimate of $\sigma$ at the center of M54 obtained by
\citet{illi} from integrated spectroscopy. \label{m54r}}
\end{figure}

\subsection{Velocity dispersion profiles}

In the upper panel of Fig.~\ref{m54r} we show the distribution of the M54 stars
in the $V_r$ vs. $r$ plane. We have divided the sample into two sets of
six and five independent radial bins of different size, respectively,
in order to keep the number of stars per bin as high as possible
while maintaining the highest degree of spatial resolution (the {\em primary} 
set of independent bins corresponds to the odd rows of Tab.~\ref{Tm54}, the
{\em secondary} set corresponds to the even rows of the table). 
The {\em primary} bins are enclosed
by the vertical lines. 
In each bin we computed the
average $V_r$ and the velocity dispersion $\sigma$, with their uncertainties
(derived with the ``jackknife'' method, as above). An iterative $3\sigma$
clipping algorithm was applied bin by bin. As we proceeded from the innermost
bin to the outer ones, any star rejected in a given bin by the clipping
algorithm was excluded from the sample. The eight rejected stars are clearly
indicated in the plot.  

The derived velocity dispersion profile is reported in the lower panel of 
Fig.~\ref{m54r} (large filled pentagons, primary set, small filled pentagons,
secondary set) and in Tab.~\ref{Tm54}. The table reports also an alternative
estimate of $\sigma$ obtained with the Gaussian maximum-likelihood method
described by \citet{walk06}, essentially equivalent to that adopted by
\citet{nicolas}. It is remarkable that in all cases the estimates obtained with
the two methods differ by much less than the reported uncertainties (in all
cases by $\le 0.3$ km/s). 
The profile is complemented with the
central estimate obtained by \citet{illi} from integrated spectroscopy of the
cluster core\footnote{Here we adopt the estimate of central velocity dispersion 
$\sigma_0=14.2$ km/s obtained by Illingworth by correcting the value 
he measured from integrated long-slit spectroscopy of the cluster center.}.
The profile shows a steep decrease from $\sigma=14.2$ km/s at the center
to $\sigma\simeq 5.0$ km/s at $r=3.5\arcmin$. Then it begins to 
grow gently up to $\sigma\simeq 9$ km/s in the last bin ($r \simeq 7\arcmin$). 

\begin{table}
\begin{center}
\caption{Velocity dispersion profile for M54. \label{Tm54}}
\begin{tabular}{cccccccc}
\tableline\tableline
$r_i$ & $r_f$ & r$_m$ & $\sigma$ & e$_{\sigma}$ & $\sigma_{ml}$& e$_{\sigma}$ & N \\
arcmin & arcmin & arcmin & km/s& km/s& km/s & km/s &   \\
\tableline
0.1& 1.0& 0.68& 10.2& 0.9&  10.1& 1.0&  95 \\
0.5& 1.5& 0.96& 10.2& 0.7&  10.1& 1.1& 131 \\ 
1.0& 2.0& 1.50&  8.6& 0.5&   8.5& 1.2& 107 \\
1.5& 2.5& 1.98&  7.4& 0.4&   7.2& 1.4& 101 \\ 
2.0& 3.0& 2.47&  7.4& 0.5&   7.3& 1.3&  86 \\
2.5& 3.5& 2.95&  6.2& 0.6&   6.1& 1.3&  66 \\ 
3.0& 4.0& 3.38&  5.3& 0.7&   5.1& 1.2&  42 \\
3.5& 5.0& 4.29&  6.3& 0.5&   6.2& 1.1&  49 \\ 
4.0& 6.0& 4.83&  6.7& 0.5&   6.6& 1.1&  53 \\
5.0& 7.0& 5.89&  7.0& 0.9&   6.9& 0.9&  36 \\ 
6.0& 9.0& 6.87&  8.8& 1.5&   8.6& 0.7&  34 \\
\tableline
\end{tabular}
\tablecomments{$r_i$ and $r_f$ are the limits of the bins, $r_m$ is the radius
of the middle of the bins. N is the number of stars in the bin, after the rejection of $3\sigma$ outliers.
$\sigma_{ml}$ is the velocity dispersion estimated with a Gaussian {\em maximum likelihood} method as done in
\citet{walk06}.}
\end{center}
\end{table}

\begin{table}
\begin{center}						      
\caption{Velocity dispersion profile for Sgr,N. \label{Tsgr}}
\begin{tabular}{cccccccc}
\tableline\tableline
$r_i$ & $r_f$ & r$_m$ & $\sigma$ & e$_{\sigma}$ & $\sigma_{ml}$& e$_{\sigma}$ & N \\
arcmin & arcmin & arcmin & km/s& km/s& km/s & km/s &   \\
\tableline
0.8& 2.0& 1.41&   9.1& 0.9&   8.9& 0.7& 41 \\
2.0& 3.0& 2.55&   8.8& 1.2&   8.6& 0.7& 38 \\
3.0& 4.0& 3.54&  10.2& 0.9&  10.0& 0.7& 56 \\
4.0& 5.0& 4.47&   8.8& 1.0&   8.5& 0.9& 43 \\
5.0& 6.0& 5.46&   9.6& 1.1&   9.3& 0.7& 40 \\
6.0& 7.0& 6.52&   8.7& 0.9&   8.5& 0.7& 43 \\
7.0& 9.0& 7.69&  10.3& 1.1&  10.1& 0.7& 56 \\ 
\tableline
\end{tabular}
\tablecomments{$r_i$ and $r_f$ are the limits of the bins, $r_m$ is the radius
of the middle of the bins. N is the number of stars in the bin, after the rejection of $3\sigma$ outliers.
$\sigma_{ml}$ is the velocity dispersion estimated with a Gaussian {\em maximum likelihood} method as done in
\citet{walk06}.}
\end{center}
\end{table}

\begin{figure}
\plotone{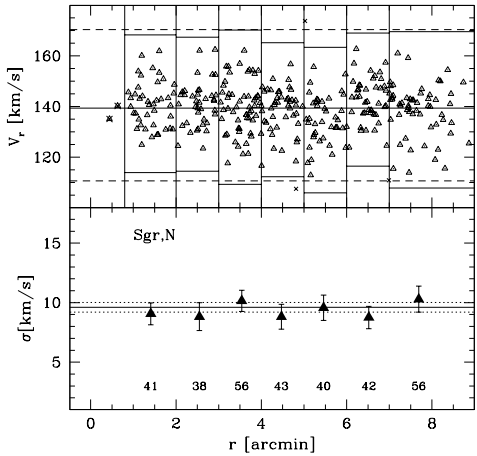}
\caption{The same as Fig.~\ref{m54r} for the Sgr,N sample (triangles replaces
pentagons here). The two innermost point are rejected for reasons of uniformity
of the sample.
In the lower panel
we also report the velocity dispersion of the whole sample as a continuous line,
$\pm$ the associated error (dotted lines).
As $\sigma$ does not appear to change with radius we use only the reported seven
independent bins.
\label{sgrr}}
\end{figure}

Fig.~\ref{sgrr} shows the velocity distribution and dispersion profile 
for the Sgr,N sample, with the same
arrangement as in Fig.~\ref{m54r}, above. The absence of any obvious trend of
velocity dispersion with radius allowed us to adopt a single set of bins of nearly 
uniform size. The dispersion in each bin has been estimated exactly as in the
case of M54, described above. 
The two innermost stars of
the sample (crosses) have been excluded from the analysis as they lie in a
too poorly sampled radial region ($r<0.8\arcmin$) and may 
affect the computation of the mean radius of the first bin. 
However, the inclusion of these stars in the
innermost bin changes the velocity dispersion estimate by $\simeq 0.2$ km/s.

It is quite obvious from the
inspection of Fig.~\ref{sgrr} that the velocity dispersion profile of Sgr,N is
completely flat over the whole radial range explored by our data. The lower
panel shows that the velocity dispersion of each radial bin is in good agreement
with the velocity dispersion of the whole sample after the clipping of
$3\sigma$ outliers, $\sigma=9.6\pm 0.4$ km/s.
The maximum observed difference among bins is $1.6\pm 1.4$ km/s; 
the maximum difference
between the dispersion in a given bin and the dispersion of the whole sample 
is $-0.9\pm 1.0$. Possibly, there may even be a weak tendency
toward lower velocity dispersions in the inner regions: in the range 
$0.0\arcmin\le r< 5.0\arcmin$ we find $\sigma=9.2\pm 0.5$ km/s (180 stars),
while in the range 
$5.0\arcmin\le r< 9.0\arcmin$ we find $\sigma=10.1\pm 0.6$ km/s (138 stars);
however this difference is clearly not statistically significant 
\cite[see][for some example of the variety of velocity dispersion profiles 
observed in dE,N]{geha,n205}. 

\begin{figure}
\plotone{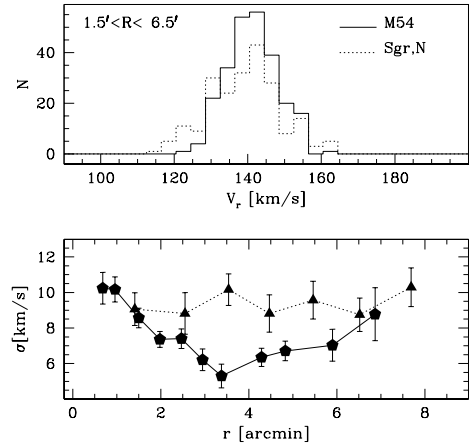}
\caption{
Upper panel: Velocity distribution of the M54 and Sgr,N samples in the radial
range in which the difference in their $\sigma$ reaches the maximum statistical
significance, according to the F test (see Tab.~\ref{Ftest}.
Lower panel: Comparison of the velocity dispersion profiles of M54 (continuous
line and pentagons) and Sgr,N (dotted line and triangles).
\label{compsi}}
\end{figure}

It is rather clear that the nucleus of Sgr and the M54 cluster have different
velocity dispersion profiles. M54 displays the typical behavior of a globular
cluster, i.e. the system becomes increasingly kinematically hot toward 
its central regions while Sgr,N has a uniformly flat profile, though the  
surface brightness profiles of the two systems are broadly similar. 
Fig.~\ref{compsi} shows the comparison
between the profiles in deeper detail.
The inner growth of $\sigma$ of M54 stars
makes the two profiles overlap at  $r \sim 1.5\arcmin$; the
innermost two points of the M54 profile are hotter than any Sgr,N point. The
outermost point of the cluster profile does match the dispersion of Sgr,N:
the velocity dispersion estimate here comes from the least populated bin as well
as the most (possibly) prone to contamination by old and metal-poor stars
belonging to Sgr (see below for a more detailed discussion). 
These are the reasons why
the comparison between the two samples 
taken as a whole does not reveal any striking
difference.
However, in the intermediate radial range  the M54
population is significantly colder than that of Sgr,N and the statistical
significance of this result is {\em very strong}.

\begin{figure}
\plotone{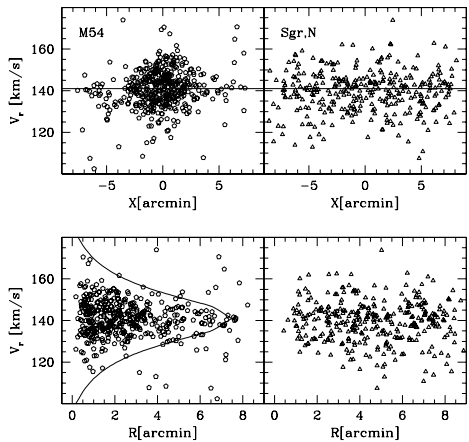}
\caption{Direct comparison of the velocity and spatial distributions of 
M54 and Sgr,N.
Upper panels: $V_r$ as a function of X for the M54 (left panel) and for
the Sgr,N (right panel) samples. 
Lower panels: $V_r$ as a function of R for the same samples. The curves in the
lower left panel are
the $\pm 3\sigma$ profiles the best-fit King model  for M54
\citep{tkd}, approximately representing the envelope of the allowed velocities 
for bound members of M54.
\label{vedif}}
\end{figure}

Tab.~\ref{KStest} reports the results of non-parametric Kolmogorov-Smirnov tests
performed on different radial ranges. In all the considered ranges the
probability that the two samples are drawn from the same parent population is
smaller than 2\% and it is smaller than 0.2\% in the 
$1.5\arcmin \le r<6.5\arcmin$ range. In the present case, however, the F
statistic is the most appropriate mean to compare statistically the two samples
\citep[$F=\sigma_{Sgr}^2/\sigma_{M54}^2$, see][and references therein]{brandt},
as the corresponding F test evaluates the probability that two samples of given
F value are extracted from Gaussian distributions {\em having the same $\sigma$}.
Tab.~\ref{Ftest} shows that in our case this probability is lower than 0.3\% in
all the considered radial range and it is as small as $<$ 0.01\% in several
intermediate ranges, including $1.5\arcmin \le r<6.5\arcmin$.
While such a large quantitative difference may not be so obviously apparent
from the dispersion profile, it is clear at a
first glance when looking at the comparisons in the X - $V_r$ and in 
the R - $V_r$ planes shown in Fig.~\ref{vedif}.  

Hence, the direct comparison of the
velocity distributions of the two samples clearly establishes that
{\em the two populations have very different kinematic properties, indicating
that the motion of M54 and Sgr,N stars is driven by different gravitational
equilibria}. 

\begin{figure}
\plotone{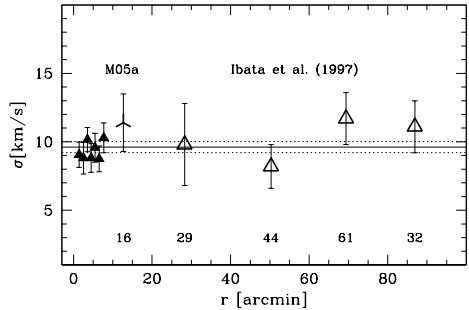}
\caption{Velocity dispersion profile of the innermost $\sim 100\arcmin$ of the
Sgr galaxy. Filled triangles are the $r<10\arcmin$ estimates obtained here, the
skeletal symbol is from the M05a stars having $[Fe/H]<-1.0$ (to avoid possible
M54 members), open triangles are from Table 2b of \citet{iba97} (CTIO-ARGUS
radial velocities). The horizontal lines are the same as in Fig.~\ref{sgrr}.
Note that this composite profile covers less than half of the core radius of the
Sgr galaxy \cite[$r_c\simeq 220\arcmin$][]{maj}.
\label{sgrtut}}
\end{figure}

Moreover, \citet{gil} clearly states that a velocity
dispersion profile monotonically declining from the center outward 
{\em ``...is an unavoidable requirement for any mass - follows
- light system...''}. In the present case, this means that while M54 behaves
like an ordinary mass - follows - light ``purely baryonic'' 
self-gravitating star cluster (at least in its innermost region, containing
most of its light/mass; see also Fig.~14 and Sect.~4.2, below), {\em 
the flat  velocity dispersion profile of Sgr,N 
implies that the mass distribution driving the kinematics of the nucleus is
significantly different from the distribution of the stars}. 
In principle, a radial gradient in the
velocity anisotropy affecting only one of the two systems may be invoked
to explane the observed  difference in the velocity dispersion profiles. As this would 
imply a correlation of the anisotropy variation with metallicity, we
regard this hypotesis as rather ``ad-hoc'' and we don't discuss it anymore, in
the following. We will re-consider the case in more detail in a future
contribution (Ibata et al., in preparation).

Finally, the velocity dispersion profile of Sgr,N does not differ significantly 
from the overall profile of the Sgr galaxy, that is found to be flat out to
large radii \citep{iba97}, as typical of dwarf spheroidals 
\cite[see][and references therein]{walk06,munoz}. 
In Fig.~\ref{sgrtut} we have combined the inner
$0.8\arcmin\le r\le 9\arcmin$ profile obtained here, 
with estimates of the dispersion at
different radii out to $r\simeq 100\arcmin$ obtained from publicly available 
data from the literature \cite[M05a,][]{iba97} 
in the same way as done for M54 and Sgr,N, above. 
Note that $r\simeq 100\arcmin$ is less
than half of the core radius of the whole Sgr galaxy \citep{maj}.
All the observations are consistent with a constant $\sigma$ in the considered
range and, in particular, there is no apparent transition in the profile at the
onset of the nuclear overdensity of stars ($r\le 10\arcmin$).
This fact strongly suggests that the inner kinematics of Sgr, including its
nucleus, are dominated by the potential set by a distribution of unseen (Dark) 
matter, not by M54 as in the scenario (b) described in Sect.~1., above.

\begin{table}
\begin{center}						      
\caption{KS test. Probability that the Sgr,N and the M54 samples
are drawn from the same parent population. \label{KStest}}
\begin{tabular}{cccccc}
\tableline
$r_i$ & $r_f$ & N$_{M54}$ & N$_{Sgr,N}$  & D$_{KS}$ & P\% \\
\tableline
0.0 & 9.0 & 417 & 318 & 1.649 & 0.87 \\ 
0.8 & 8.0 & 353 & 300 & 1.531 & 1.84 \\ 
2.0 & 7.0 & 199 & 219 & 1.624 & 1.02 \\ 
2.0 & 6.0 & 181 & 177 & 1.698 & 0.63 \\
1.5 & 6.5 & 247 & 213 & 1.817 & 0.13 \\
\tableline
\end{tabular}
\tablecomments{KS = Kolmogorov-Smirnov.
$r_i$ and $r_f$ are the limits of the considered radial ranges,
N$_{sample}$ is the number of stars of the sample in the considered radial
range. $3\sigma$ outliers have been excluded from the samples.}
\end{center}
\end{table}
\begin{table}
\begin{center}						      
\caption{F test. Probability that the Sgr,N and the M54 samples
are drawn from Gaussian distributions having the same $\sigma$. \label{Ftest}}
\begin{tabular}{cccccc}
\tableline
$r_i$ & $r_f$ & N$_{M54}$ & N$_{Sgr,N}$  & F & P\% \\
\tableline
 0.0 & 9.0 & 417 & 318 & 1.345 &   0.23 \\ 
 0.8 & 8.0 & 353 & 300 & 1.448 &   0.04 \\  
 2.0 & 7.0 & 199 & 219 & 2.010 &  $<$0.01 \\
 2.0 & 6.0 & 181 & 177 & 2.019 &  $<$0.01 \\  
 1.5 & 6.5 & 247 & 213 & 1.978 &  $<$0.01 \\
 2.5 & 3.5 &  65 &  51 & 2.546 &   0.02 \\
 1.5 & 3.5 & 166 &  80 & 1.862 &   0.04 \\
\tableline
\end{tabular}
\tablecomments{$r_i$ and $r_f$ are the limits of the considered radial ranges,
N$_{sample}$ is the number of stars of the sample in the considered radial
range. $3\sigma$ outliers have been excluded from the samples.
F is computed as $F=\sigma_{Sgr,N}^2/\sigma_{M54}^2$.}
\end{center}
\end{table}

\begin{figure}
\plotone{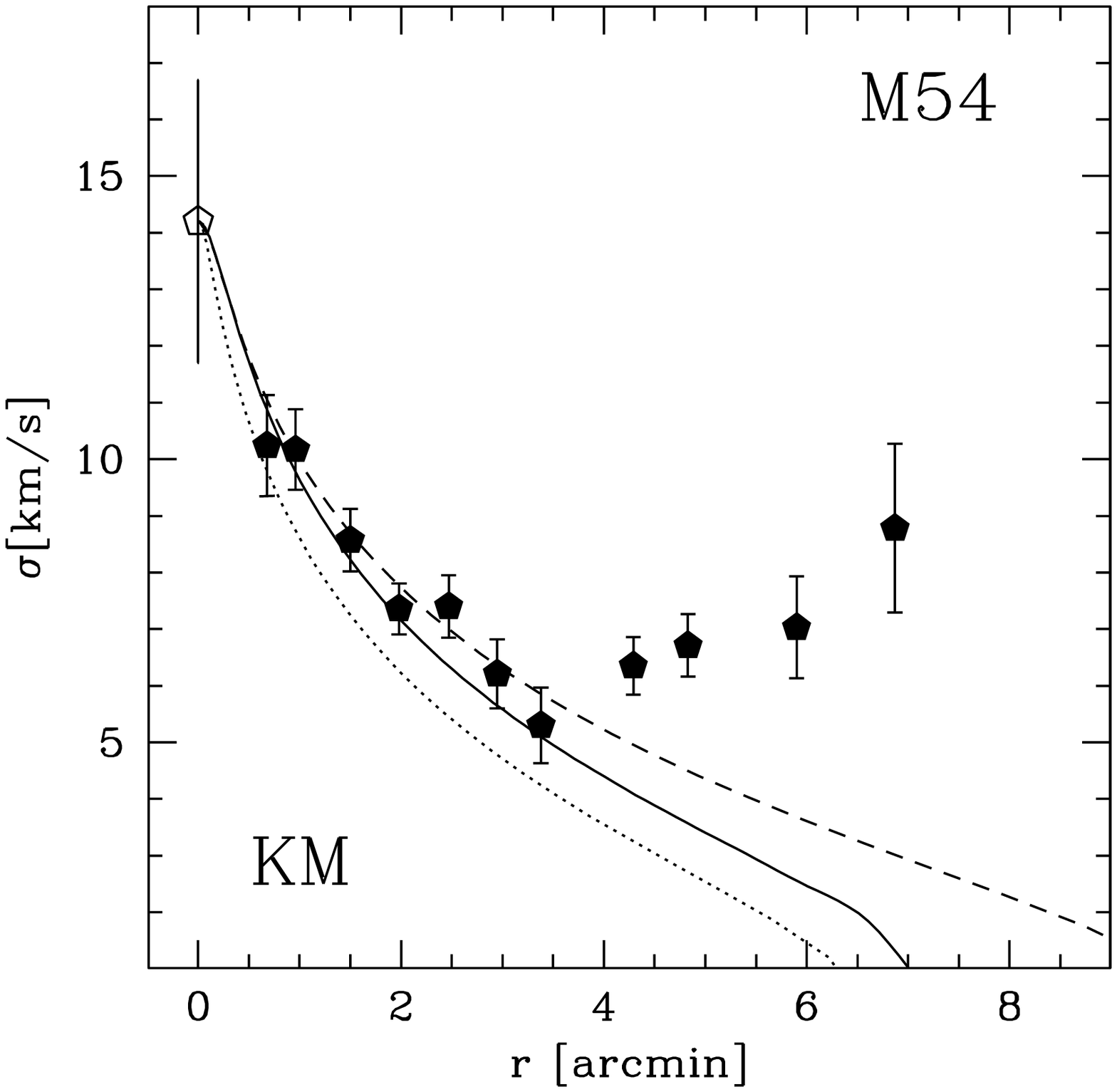}
\caption{Comparison of the observed velocity dispersion profile of M54 with
King models that best fit the Surface Brightness profile of the cluster, 
as proposed by various authors (see Tab.~\ref{m54dat}), and
normalized to $\sigma_0=14.2$ km/s. 
Continuous line: $r_c=0.11\arcmin$ and $C=1.85$, from \citet{tkd}. 
Dotted line: 
$r_c=0.053\arcmin$ from \cite{ng6} and $C=2.10$, obtained by adopting the 
estimate of
the tidal radius from Trager et al., since it is not provided by Noyola \&
Gebhardt. Dashed line: $r_c=0.09\arcmin$ and $C=2.05$, from \citet{mcv}.
\label{siking}}
\end{figure}

\subsection{Comparison with models}

In Fig.~\ref{siking} the observed velocity dispersion profile of M54 is compared
with the theoretical profiles of single-mass isotropic
\citet{king} models that have been proposed 
by various authors as best-fits to the surface brightness profile of the cluster
(see Tab.~\ref{m54dat}). The models proposed by \citet[][continuous line]{tkd}
and by \citet[][dashed line]{mcv} provide a reasonable description of the data
out to $r\la 4\arcmin$, while there is a clear tendency of the data to be
hotter than the models for $r>4\arcmin$. The outermost observed point,
at $r\simeq 7\arcmin\la r_t$, displays a velocity dispersion significantly
higher than what is predicted even by the hottest model 
$[\sigma_{obs}-\sigma_{KM}]_{r\sim 7\arcmin}=5.8\pm 2.0$ km/s.
There are two possible explanations for this unexpected rise of the
velocity dispersion curve near the tidal radius of the cluster. 

First, while our
selection criteria prevents any contamination of the Sgr,N sample by M54 stars, 
since cluster stars with $[Fe/H]\ge -0.8$ simply do not exist, it is well
known that the Sgr galaxy does host a minority of stars with $[Fe/H]\le -1.0$
\citep[$\la $12\%,][]{lbhb}, hence some degree of contamination of the M54
sample by Sgr stars is expected (see Sect.~3.1) and the less populated outer
bins of the M54 profile are clearly the most sensitive to the effect of the
inclusion of contaminants. Rough estimates indicate that the contamination by
metal poor Sgr stars should amount to less than 3\% in the innermost $2\arcmin$,
less than 10\% for $2\arcmin\le r<4\arcmin$, but it can raise to $\la$ 20\%
for $4\arcmin\le r<6\arcmin$ and be even larger for $r>6\arcmin$.
So, contamination by metal-poor Sgr stars seems a viable explanation for the
raise of the dispersion profile of M54 such that it becomes similar 
to that of the Sgr,N sample in the outermost, and presumably most contaminated,
bin.

Alternatively, it may be conceived that, during its spiraling toward the center
of Sgr,N (see Sect.~5.2, below), M54 suffered some tidal harassment from the
host galaxy (Sgr). In this case, some stars in the last bins would have been
stripped from the outskirts of the cluster and heated up to the same 
dispersion of the surrounding field \cite[see][and references therein]{munoz}. 
In the lower left panel of Fig.~\ref{vedif}
the $\pm 3\sigma$ dispersion profiles of the best-fitting King model for M54
\citep{tkd} are superposed on the M54 sample in the R vs. $V_r$ plane: the good
match between the shape of the bulk of the observed distribution and the model
profile out to the tidal radius strongly suggests that most of the velocity
outliers at large radii are likely not gravitationally bound to the cluster, 
independent of their origin (unrelated metal poor Sgr stars or former M54
members that have been tidally stripped).

While we consider the ``contamination'' hypothesis as more likely, 
the ``tidal'' hypothesis is
a very fascinating possibility and it cannot be dismissed with the data we 
have presently in hand (see also Sect.~ 4.3, below). 

Turning back to the inner regions of the cluster, the main
conclusion we can draw here is that 
the observed velocity dispersion profile is consistent with the 
expected kinematics of the King models that best fit the light distribution 
of the cluster. Hence, M54 has the kinematics of an ordinary  
{\em self-gravitating}  globular cluster, at least in its innermost 
$\sim 25$ pc, enclosing more than 90\% of the cluster light/mass.

\begin{figure}
\plotone{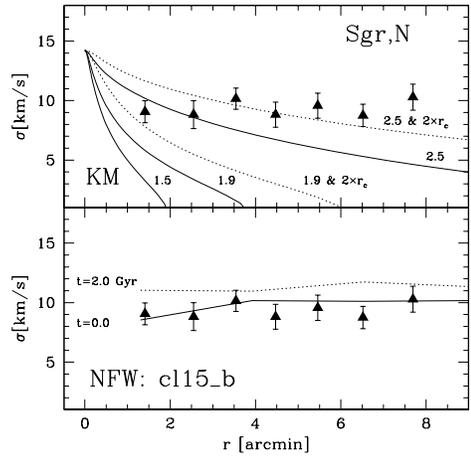}
\caption{Comparison of the observed velocity dispersion profile of
Sgr,N with theoretical models. Upper panel: comparison with King models of
$r_c=0.05\arcmin$ (best fit value) and various C (thin continuous lines labeled with their
C value), normalized at the central velocity dispersion of M54 ($\sigma_0=14.2$
km/s). The $C\sim 1.9$ case corresponds to the model that best-fits the observed
SB profile and is plotted with a thicker line. 
The dotted curves show the effect on the profile of the assumption of $r_c$ and
$r_t$ values much larger than the best fit value (of a factor $\times 2$).  
Lower panel: the observed velocity dispersion profile is compared with the
profile of the NFW model adopted in the N-body simulation cl15\_b 
(see Tab.~\ref{Tnbody}). The profiles at the beginning (continuous line) and at
the end (dotted line) of the simulation are shown.  
\label{sisag}}
\end{figure}

\subsubsection{The case of Sgr,N}

In Fig.~\ref{sisag} the observed velocity dispersion profile of Sgr,N is
compared with various theoretical models.
In the upper panel we compare King's models with $r_c=0.05\arcmin$ 
and various values of the concentration parameter C (the best-fit model 
for the SB profile corresponds to C=1.9). All the theoretical
profiles  have been normalized at the central
velocity dispersion of M54: this is a quite arbitrary choice, the family of
profiles can be shifted up and down according to the preference of the reader.
It is clear that, independent of the adopted normalization, none of the
plotted profiles provides a satisfactory match to the observations. 
The dotted profiles, reported here for comparison, show the effects of 
factor $2\times$ changes in the adopted values of $r_c$ and $r_t$. Even if such
large values were compatible with the observations presented in Sect.~2,
above (and they are not), the corresponding profiles remain unable to 
reproduce the observed dispersions.

\begin{figure*}
\plottwo{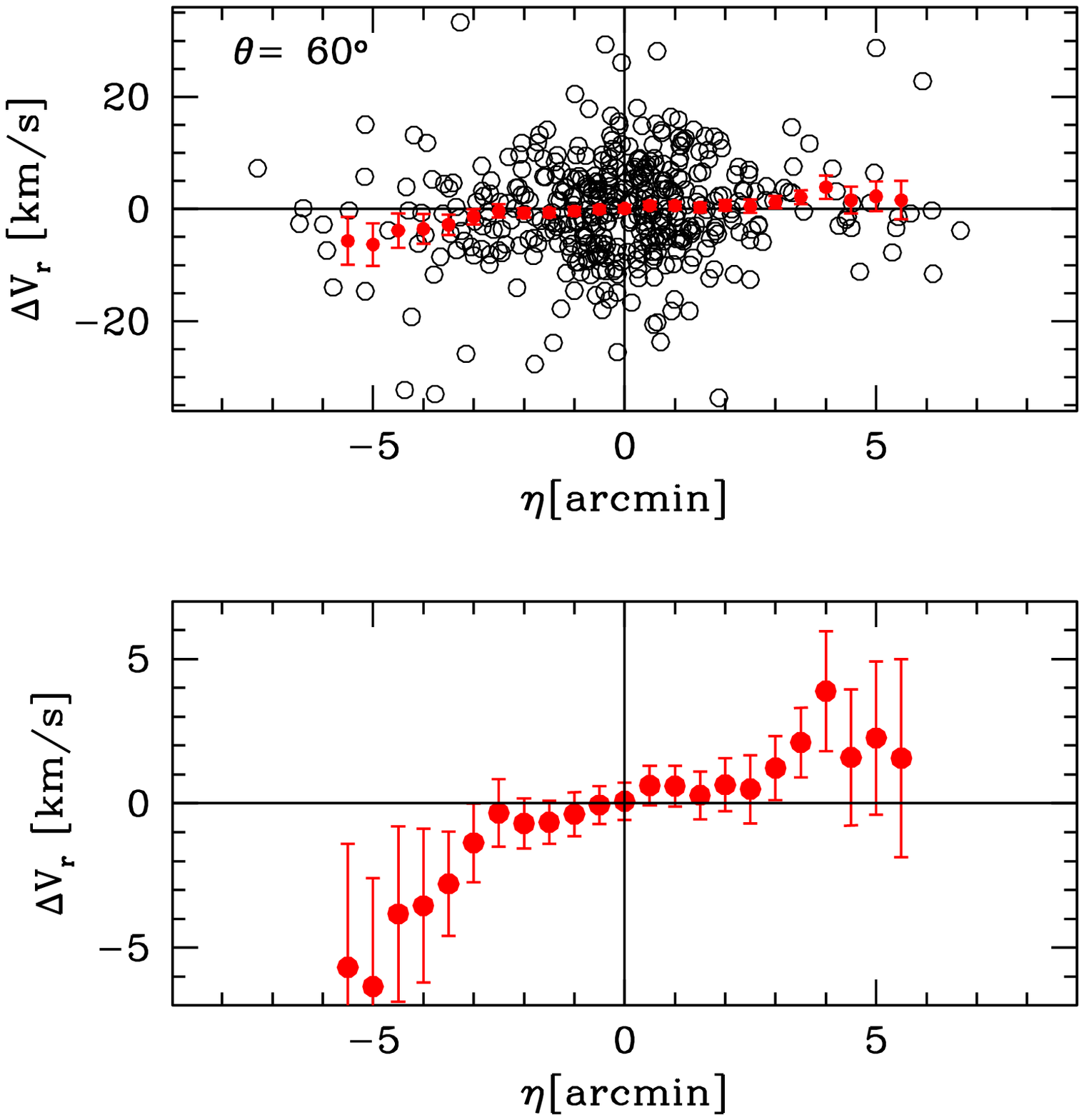}{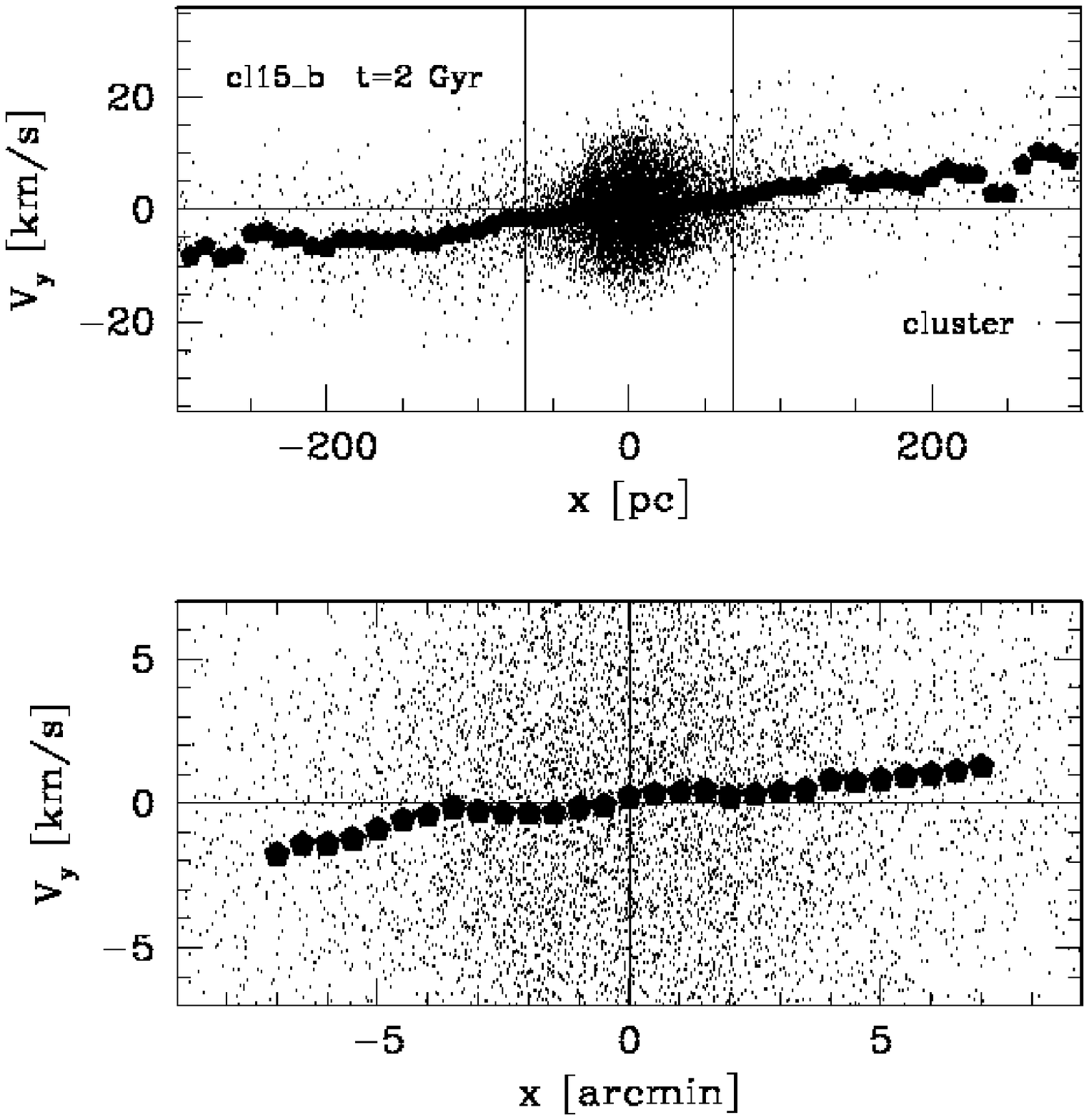}
\caption{{\em Left Panels:} 
Rotation curve of M54 assuming a rotation axis tilted by 
$\theta=60\degr$ East of North.
$\Delta V_r$ is the difference between the observed radial velocity and the 
global mean systemic velocity of Sgr,N ($\langle V_r\rangle = $ km/s).
Upper panel: individual stars. Lower panel: running mean with bin width of
$2^\prime$ and step of $0\mcnd5$.
{\em Right Panels:} Rotation in the remnant of the model of M54 at the end of the
simulation cl15\_b. The coordinate system is chosen such that x,z projects on
the plane of the sky while y is along the line of sight and the orbit of the
cluster is in the x,y plane. 
Upper panel: l.o.s. velocity as a function of x for
the innermost 300 pc. The vertical thin lines enclose a region of the same
dimension as the one studied here; the large dots are the running mean of the
velocity computed with the same bin width and step as in the real case.
Note
the tidal tails emerging from the main body of the remnant at $x \sim \pm 50$
pc. Lower panel: zoomed view of the inner region of the plot. In this case x has
been converted in arcmin by placing the remnant at the same distance as the real
cluster. The diagram has the same scale as the lower left panel 
for an easy comparison.
\label{rotaz}}
\end{figure*}

One could be tempted to suggest that any King model with
sufficiently large scale to fit the whole surface profile of the Sgr galaxy
\citep[as the C=0.90, $r_c=224\arcmin$ model by][]{maj} would have a nearly flat
dispersion profile on the relatively small radial range studied here ($r\le
9\arcmin$), thus providing a reasonable fit to the observed profile. 
This is true, but these models
would completely fail in reproducing the observed Surface Brightness profile
in the same range {\em by more than} 10 mag/arcsec$^2$ (see Fig.~\ref{prprof},
above), since they lack any nuclear overdensity at their center. 
 
Hence, contrary to the case of M54, King models that fit the
surface brightness profile of Sgr,N appear unable to reproduce its kinematics:
this further supports the conclusion that M54 and Sgr,N are systems of very
different nature\footnote{We stress here that we are not attaching any
particular physical meaning to King models while fitting Sgr,N profiles.
This is just another way to put in evidence the
differences between Sgr,N and M54 and a convenient reference model to
parametrize the observed SB profile \citep{munoz}.}. 
In particular, as already noted, mass - follows -light models
are incompatible with the observed velocity dispersion profile of Sgr,N
\citep{gil}.

In the lower panel of
Fig.~\ref{sisag} the observed profile is compared to those obtained from
the N-body realization of a \citet[][hereafter NFW]{nfw} model of the suite that
is described in Sect.~5, below (model NFW3, Tab.~\ref{Tnbody}).
The continuous line is the inner velocity dispersion profile of the NFW3 model
at the beginning of the simulation; the dotted line is the profile at the end of
the simulation, i.e. after the complete orbital decay of a model of M54 that is
launched in orbit within the NFW3 halo. 
It is interesting to note that the
theoretical profiles are fairly flat in the considered radial range, 
in good agreement with the data. It may be conceived that the (baryonic) nucleus
of Sgr lies within the inner cusp of the NFW halo of dark matter that is
presumed to embed the Sgr galaxy \citep[][and references therein]{iba98,maj}.
The potential in this inner region would be dominated by the DM cusp that will
impose the flat dispersion curve to the observed Sgr,N stars. It does not seem
necessary to assume a strict correlation between the typical size and/or density
profile of the DM cusp/core and those of the embedded stellar nucleus: the DM
overdensity would have simply provided the ``local'' minimum of the
overall potential well to ``attract'' the largest density of the infalling gas
that formed Sgr,N. It may be interesting to note that the NFW3 model, 
whose velocity dispersion profile is shown in the lower panel of 
Fig.~ref{sisag}, encloses a mass of $3.2\times 10^6 M_{\sun}$ within the range
of projected radii covered by our data ($r_p\simeq 70$ pc), and that the
region in which its projected density profile is well fitted by a  power-law,
i.e. the inner cusp, has a size of $r_p\la 100$ pc, similar to Sgr,N.

Unfortunately, we have no conclusive observational evidence in support of the
above scenario. Note also that a DM halo with a core much larger than the
nuclear scale would also display a flat dispersion curve, in this region (as
discussed above for the case of the C=0.90, $r_c=224\arcmin$ King model); it is
just the small-scale inner cusp as a possible seed for the formation of a 
barionic nucleus that can make the NFW option slightly more attractive.

A flat velocity dispersion profile may arise as a consequence of strong tidal
disturbance, in mass-follow-light models of dwarf galaxies \citep{munoz}. While
Sgr is clearly in course of tidal disruption, here we are dealing with the
kinematics of the innermost $\simeq 70$ pc of the galaxy, that should 
(reasonably)
be considered as untouched by the Galactic tides, since the tidal radius of Sgr 
is as large as a few thousands of parsecs  
\cite[the formal King's major
axis limiting radius is $r_t\ga 10$ kpc, according to][]{maj}.

Finally, it may appear as a curious coincidence that M54 and Sgr,N have similar
spatial scales, if they have independent origin. In this sense we can only
note that the distributions of sizes and luminosities of globular clusters and
stellar galactic nuclei largely overlap, so the observed similarity may not be
particularly odd \cite[see][]{lucky,bok07}. Moreover, it should be noted that
the core and half-light radii of M54 are $\sim 2$ times larger than those of
Sgr,N, clearly a non-negligible difference. Finally, while the best fitting King
models have essentially the same C parameter, 
the two observed profiles appear to
have significantly different slopes at large radii, where the profile of Sgr,N
departs from the King model.

\subsection{Rotation}

We searched for signal of rotation in the two samples, also taking into account
the correction for perspective rotation, according to \citet{ven} and taking the
proper motion of Sgr/M54 from \citet{dinescu}; in the present case the 
absolute value of the maximum correction is $\le 1.0$ km/s. 
Given a rotation of the coordinate axes by an angle $\theta$
\begin{displaymath}
\eta=X cos(\theta)-Y sin(\theta)
\end{displaymath}
\begin{displaymath}
\chi=X sin(\theta)+Y cos(\theta)
\end{displaymath}
we tried all the possible values of $\theta$ in steps of $1\degr$. For each
adopted $\theta$ we computed the median velocity of stars with negative and
positive $\eta$. The difference between the two median velocities  is a robust
measure of the amplitude of any systemic motion around an axis tilted 
by $\theta$
degrees from the $\chi$ axis, i.e. of the rotation of stars about this axis.

In some cases, ordered patterns of $V_r$ as a function of $\eta$ emerged
(in particular for the M54 sample), but
their amplitude was quite weak, typically lower than 2.0 km/s. While we can
clearly exclude the presence of rotation signals stronger than this, we cannot
exclude that such weak patterns are in fact real.  
In Fig.~\ref{rotaz} we show the best case for the M54 sample
compared with the end product of one of the N-body simulations that are
described in Sect.~5, below. We refrain from attaching any significance to such a
weak pattern, in particular as it has an amplitude similar to the typical
uncertainty of the individual velocity measures and as it is much weaker than the
typical velocity dispersion of the sample. We just want to draw the attention of
the reader to the fact that the orbital decay of a massive cluster within the
parent dwarf galaxy seems to naturally produce these kinds of weak 
velocity gradients in the inner part of the sinking cluster, at the end of the
decay process 
\cite[see][for a deeper discussion of the effects of tidal disruption]{munoz}.
However, it is quite clear that the detection at $r\ga 10\arcmin$ of stars 
having metallicity compatible with M54, possibly lower velocity dispersion 
than the surrounding metal-rich Sgr population, and, above all, having
significant differences in mean systemic velocity compatible with a tidally
induced rotation like that shown in the upper right panel of Fig.~\ref{rotaz}, 
would be the final ``smoking-gun'' of the orbital decay of M54 to its current
position. On the other hand the failure to find such a component would not
be sufficient to rule out the hypothesis as, for instance, the density of stars
in the tidal tails is expected to be {\em very} low and the ``rotation'' signal
may be greatly weakened by unfavorable orientations of the orbital plane with
respect to the line of sight \citep{munoz}.

\section{The kinematics of M54 and Sgr,N: N-body simulations}

M05a used simple analytical formulae to verify the plausibility of the
hypothesis that the ``exact''  coincidence between the positions of M54 and
Sgr,N is due to the cluster progressively spiraling into the center of density of
the Sgr galaxy due to dynamical friction. 
They found that the observed status of
the globular cluster system of Sgr is in full agreement with this hypothesis.
The orbit of the massive M54 cluster is expected to decay completely within one
Hubble time if it is born within 3 core radii ($\sim 5$ kpc) from the center of
Sgr, while the other, much less massive Sgr clusters 
(Ter~7, Ter~8, Arp~2) would be
essentially unaffected by dynamical friction if they were born outside 1 core
radius. While promising, this first result demands further investigation. One
very interesting question, for instance, is: does the actual orbital decay of M54
under realistic conditions lead to such a nearly perfect coincidence 
between M54 and
Sgr, N as observed? To answer questions like this we performed the suite of 
N-body simulations that is described in detail below. While examining the
results of our N-body experiments it should be carefully considered
that:     

\begin{itemize}

\item The main purpose of our simulations is to study the effect of dynamical
friction in the specific case of M54 orbiting within the Sgr galaxy, with
particular focus on the final status of the system. 
Detailed and more general 
theoretical analyses of the effects of dynamical friction on globular
clusters within a dwarf galaxy can be found in \citet{oh}, \citet{read},
\citet{fujii}, \citet{mio06}, and
\citet{sanchez}, and references therein.

\item We do not intend our simulations to be exhaustive of all the possible cases.
This would be prohibitive since, for example, we do not even know what
is the total mass of the bound part of the Sgr galaxy 
\cite[see][and references therein]{maj}. Our aim is to verify if a broadly
realistic case can produce the final outcome we actually observe, i.e. a
cluster that has nearly exactly the same position and {\em velocity} as 
the nucleus of its host galaxy.

\item Here we study the case of a massive cluster orbiting within a model of 
the
Sgr galaxy which evolves in isolation. We verified that this assumption does not
seriously affect our main results. We replicated some of our experiments
launching the Sgr model into a realistic (but static) Galactic potential 
(model 2d of \citealt{dehnen98}): we found that if M54 was not lost into the tidal tails of the
disrupting Sgr galaxy \citep[as it happened for some Sgr clusters, see][]{mic3}, 
it plunged to the center of Sgr approximately on the same timescales as the 
isolated models.

\item A detailed comparison with the existing literature on dynamical friction
would be clearly beyond the scope of the present paper 
\cite[see, for instance][and references therein]{colpi}. 
In the present context,
suffice to say that our results are in reasonably good agreement with the
analytic estimates by M05a as well as with the  results by
\citet[][and references therein]{oh,sanchez,read}.

\end{itemize}

\subsection{General features of the N-body experiments} 

The simulations were performed with {\tt falcON}, a fast and momentum-conserving
tree-code \citep{den00,den02}, within the NEMO environment \citep{nemo}. 
Gravity was softened with the kernel `P$_2$' \citep{den00}, with a softening
length of 3 pc (except for two special cases, bh\_15p and bh\_02a, discussed
in Sect.~5.2, below). The minimum time-step was 
$1.9\times 10^{-6}$ Gyr, and
the tolerance parameter $\theta=0.6$.

To model the Sgr galaxy we adopted a truncated \citet[][hereafter NFW]{nfw} model 
of the form:
$$\rho(r) \propto {{{\rm sech}(r/b)}\over{r (r+a)^2}} \, .$$
For the model NFW1, the parameters $a=0.1$~kpc, $b=2.5$~kpc, were selected, which
together with a maximum circular speed of $V_c=30$~km/s, gives a total
mass of $M_{TOT}=2.4 \times 10^8 M_{\sun}$. This rather massive model was chosen
to approximate to the sort of dense system suggested by \citet{iba98} that would 
not be rapidly destroyed by Galactic tides.
A second, lower mass system (NFW2) with $a=0.1$~kpc, $b=2.5$~kpc and $V_c=15$~km/s
($M_{TOT}=6.1 \times 10^7 M_{\sun}$) was also simulated. The motivation for this
second model was that the central velocity dispersion in this case should be similar
to the observed dispersion of Sgr 
(while keeping $a$ and $b$ identical to the first model). The NFW3 
with $a=0.5$~kpc, $b=15.0$~kpc and $V_c=17$~km/s was chosen to have
both a size and a central velocity dispersion similar to the present-day 
main body of Sgr, irrespective of its robustness to Galactic tides. 

In all cases the NFW halos were modeled with 10$^5$ particles.
M54 was modeled as a single massive particle of 1, 1.5 or 2 million solar masses,
as if it were a black hole ({\em bh} simulations, as opposed to {\em cl} 
simulations, see Tab.~\ref{Tnbody} and Sect.~5.2, below). 
The particle is launched within the NFW halo from 
(x,y,z)=(2,0,0) kpc or (4,0,0) kpc\footnote{In the coordinate system in which
(x,y,z)=(0,0,0) kpc corresponds to the center of density of the model halo and
a point having $(V_x,V_y,V_z)=(0,0,0)$ km/s is at rest with respect to 
the center of the system.
 The adopted range of initial distances from the center of the
parent galaxy is typical of the other clusters residing in the main body of Sgr
(Ter~7, Arp~2, Ter~8) which are located between $\sim 3$ kpc and $\sim 5$ kpc
from the center of Sgr and should have orbits that are stable against dynamical
friction (see M05a and below).} 
\citep[note that the core radius of Sgr is $r_c\simeq 1.7$ kpc][]{maj}, with
velocities in the y direction in the range 4.0 km/s $\le V_y\le$  12.0 km/s, while 
$V_x=V_y=0.0$ km/s, hence the orbits of the M54 models lie in the x,y plane. 
As a mere convention to simplify the discussion, we consider
the y direction coincident with  the line-of-sight between us and the center of
Sgr, and the x and z directions as projected on the plane of the sky.
When needed we will select particles according to their projected radius
$r_p=\sqrt{x^2+z^2}$. 

\begin{figure*}
\plottwo{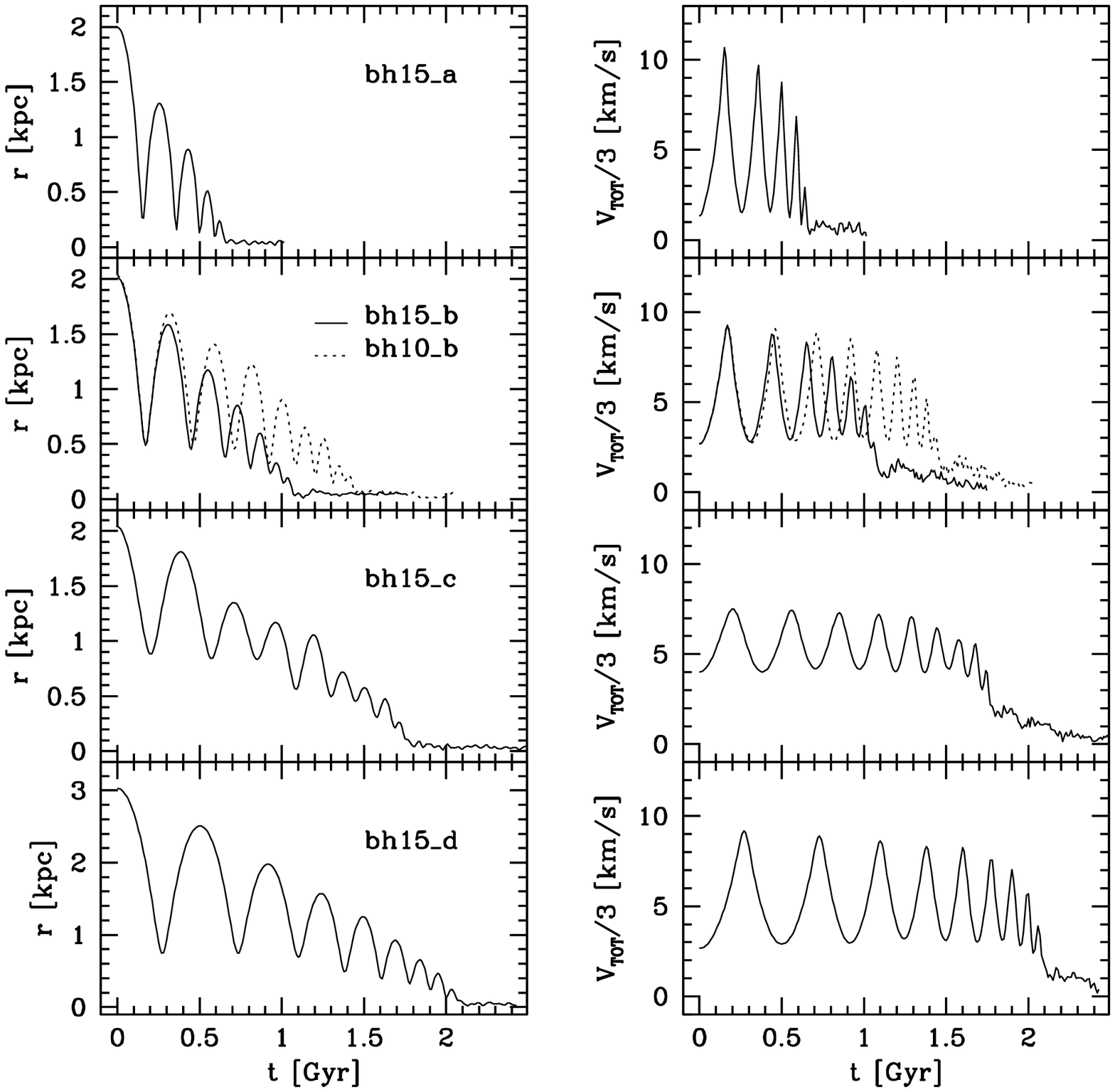}{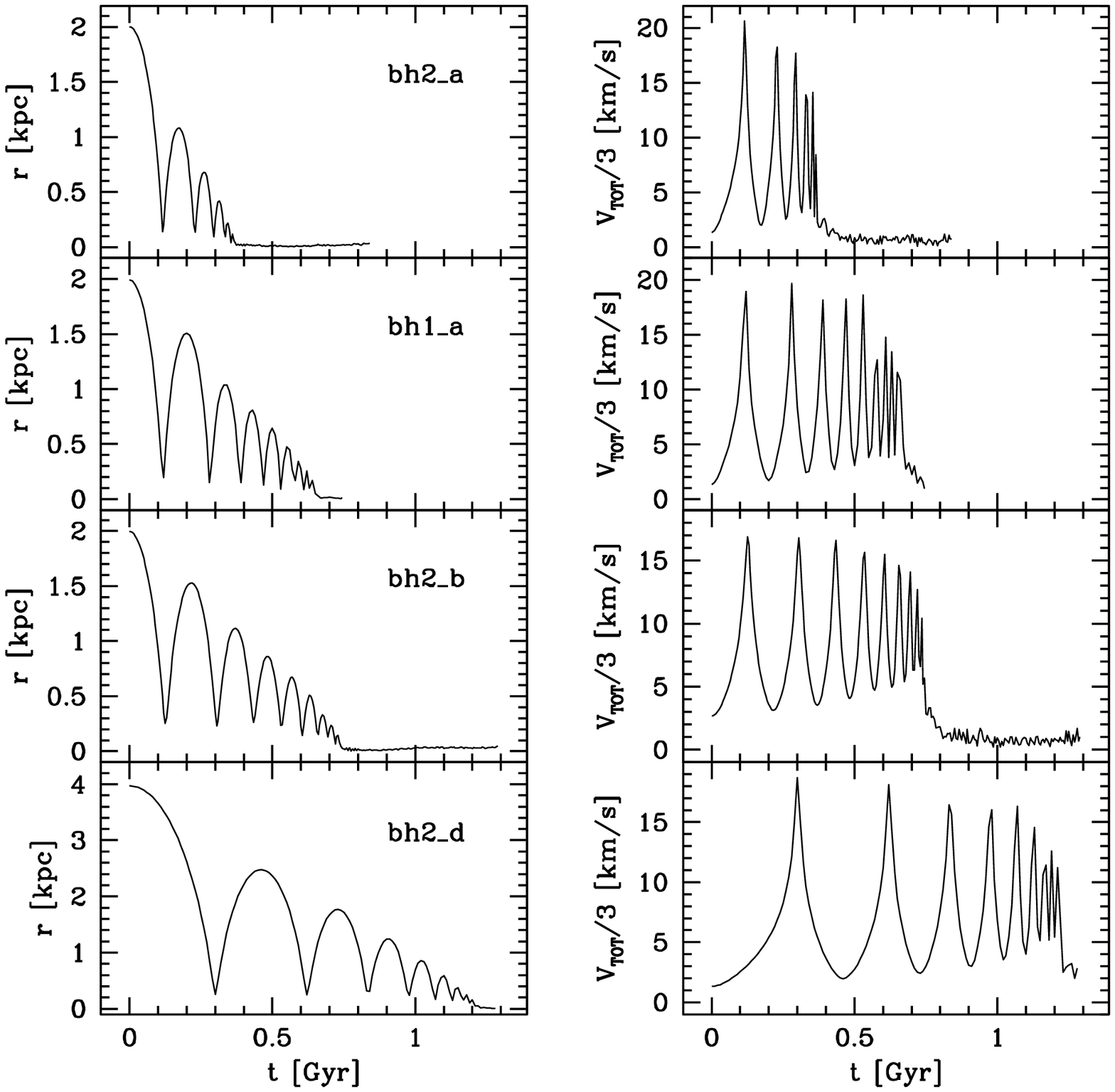}
\caption{Orbital evolution of the single massive particle representing M54 in
the N-body experiments described in Tab.~\ref{Tnbody}. {\em Left panels}:
distance from the center of the host NFW halo as a function of time.
{\em Right panels}: mean one-component velocity of the massive particle 
\slantfrac{V_T}{3} as a function of time, 
where $V_T=(V_x^2+V_y^2+V_z^2)^{\slantfrac{1}{2}}$.
\label{orb2}}
\end{figure*}

The initial conditions of the various simulations are reported in 
Tab.~\ref{Tnbody}. The simulations were usually stopped when it was clear that
the orbit of the M54 particle had reached a stable configuration 
at the center of the host halo(t$_{tot}$).
The {\em decay time} (t$_{d}$) is defined as the epoch at which the 
3-D apocentric distance of
the M54 particle become $r_{apo}\le 30$ pc. Note that around this limit the adopted
time step becomes too small with respect to the orbital period of the infalling
massive particle, hence the evolution of the orbit cannot be followed further. 
To provide a quantitative idea of the
evolution of the eccentricity of the orbit 
($\epsilon=$ ($r_{apo}-r_{peri}$)/($r_{apo}+r_{peri}$)) we report the
eccentricity at the first ($\epsilon_1$) and fifth ($\epsilon_5$) pericentric
passages. The evolution of the orbital radius and of the average velocity 
of the M54 models is shown in Fig.~\ref{orb2}. 

\begin{table*}
\begin{center}
\caption{N-body simulations: fundamental parameters and results.\label{Tnbody}}
\begin{tabular}{lccccccccccc}
\tableline\tableline
name & halo & M$_{halo}$ & N$_{halo}$ & M$_{clus}$& N$_{clus}$&X$_0$ & V$_{Y,0}$ & t$_d$ & t$_{tot}$ & $\epsilon_{1}$ & $\epsilon_{5}$\\ 
     &      & $M_{\sun}$ &            & $M_{\sun}$& & kpc   & km/s  &  Gyr  &   Gyr   &		&		\\   
\tableline
bh2\_a  & NFW1 & $2.4\times 10^8$  &$10^5$& 2.0$\times 10^6$  & 1      &2.0 & 4.0 & 0.37 & 0.84 & 0.86 & 0.67\\
bh2\_b  & NFW1 & $2.4\times 10^8$  &$10^5$& 2.0$\times 10^6$  & 1      &2.0 & 8.0 & 0.74 & 1.28 & 0.73 & 0.52\\
bh2\_c  & NFW1 & $2.4\times 10^8$  &$10^5$& 2.0$\times 10^6$  & 1      &2.0 &12.0 & 1.28 & 1.58 & 0.56 & 0.45\\
bh2\_d  & NFW1 & $2.4\times 10^8$  &$10^5$& 2.0$\times 10^6$  & 1      &4.0 & 4.0 & 1.22 & 1.28 & 0.87 & 0.65\\
bh2\_e  & NFW1 & $2.4\times 10^8$  &$10^5$& 2.0$\times 10^6$  & 1      &4.0 & 8.0 & 2.90 & 3.00 & 0.68 & 0.65\\
bh1\_a  & NFW1 & $2.4\times 10^8$  &$10^5$& 1.0$\times 10^6$  & 1      &2.0 & 4.0 & 0.67 & 0.74 & 0.83 & 0.72\\
bh1\_b  & NFW1 & $2.4\times 10^8$  &$10^5$& 1.0$\times 10^6$  & 1      &2.0 & 8.0 & 1.19 & 1.34 & 0.72 & 0.58\\
\tableline
bh1\_N  & NFW2 & $6.1\times 10^7$  &$10^5$& 1.0$\times 10^6$  & 1        &2.0 & 4.0 & 2.25 & 2.50 & 0.36 & 0.13\\
cl1.2\_N& NFW2 & $6.1\times 10^7$  &$10^5$& 1.2$\times 10^6$& $10^4$   &2.0 & 4.0 & \nodata & 2.50 & \nodata & \nodata\\
\tableline
bh15\_a & NFW3& $4.2\times 10^8$  &$10^5$& 1.5$\times 10^6$  & 1       &2.0 & 4.0 & 0.80 & 1.00 & 0.76&  0.71 \\
bh15\_b & NFW3& $4.2\times 10^8$  &$10^5$& 1.5$\times 10^6$  & 1       &2.0 & 8.0 & 1.25 & 1.70 & 0.61&  0.44 \\
bh15\_c & NFW3& $4.2\times 10^8$  &$10^5$& 1.5$\times 10^6$  & 1       &2.0 &12.0 & 2.05 & 3.00 & 0.40&  0.36 \\
bh15\_d & NFW3& $4.2\times 10^8$  &$10^5$& 1.5$\times 10^6$  & 1       &3.0 & 8.0 & 2.10 & 2.42 & 0.60&  0.52 \\
cl15\_b & NFW3& $4.2\times 10^8$  &$10^5$& 1.5$\times 10^6$  & $10^4$  &2.0 & 8.0 & \nodata & 2.00 &  \nodata&  \nodata  \\
bh10\_b & NFW3& $4.2\times 10^8$  &$10^5$& 1.0$\times 10^6$  & 1       &2.0 & 8.0 & 1.60 & 2.05 & 0.61&  0.49 \\
\tableline
bh15\_p & NFW3& $4.2\times 10^8$  &$10^5$& 1.5$\times 10^6$  & 1       &5.0 &12.0 &--$^{*}$  & 6.00 & 0.25&  0.15 \\
bh02\_a & NFW3& $4.2\times 10^8$  &$10^5$& 2.0$\times 10^4$  & 1       &2.0 & 4.0 &--$^{**}$ & 6.00 & 0.75&  0.75 \\
\tableline
\end{tabular}
\tablecomments{NFW1: $V_c=30$ km/s; NFW2: $V_c=15$ km/s. 
NFW1, NFW2: $a=0.1$ and $b=2.5$. NFW3: $V_c=17$ km/s; $a=0.5$ and $b=15.0$.
t$_d$: time at which the orbit is decayed to
a distance from the center $r\le 30$ pc. t$_{tot}$: total time of the
simulation. cl1.2\_N: t$_d$ cannot be defined as in the above cases since the cluster is no more self-bound at
$t\ga 0.8$ Gyr; at $t\sim 1.0$ the remnant is approximately at the center of 
the host halo. The case of cl15\_b is analogous.\\ 
$^{*}$ At the end of the simulation the BH has reached a distance of $\sim 2$ kpc from the center of the halo.\\
$^{*}$ The orbit of the BH remain stable over the whole duration of the simulations.\\
}
\end{center}
\end{table*}

\subsection{The evolution of the orbit of M54 models} 

From the inspection of Tab.~\ref{Tnbody} and Fig.~\ref{orb2} we can draw the
following conclusions:

\begin{enumerate}

\item In spite of the wide range of initial conditions that has been explored,
the orbit of all the M54 models decayed completely in less than 3 Gyr 
(5-15 orbits).
At the end of our simulations M54 is always virtually {\em at rest} (mean
velocity $\la 1-2$ km/s) at the very center of the host halo (within a few
softening lengths). Hence the hypothesis that M54 reached its presently observed
status by dynamical-friction-driven orbital decay appears completely realistic
and viable (within the limitations imposed by the resolution of our N-body
experiments).

\item Two special simulations (bh15\_p and bh02\_a) have been performed at a
lower resolution (softening length of 10 pc) as the BH was not expected to reach
the densest regions of the host halo (NFW3). bh15\_p explores the case of 
a M54-like point mass decaying from somehow extreme initial conditions (large
distance - $X_0=5.0$ kpc - and high velocity $V_{Y,0}=12.0$ km/s). In six Gyr
the orbit of the BH is sufficiently decayed to bring it within 2 kpc from the
center of the halo. The other simulations indicates that from this point it will
require less than 3 Gyr to reach the very center of the halo. From the bh15\_p
simulation we can deduce that for starting distances larger than 5.0 Kpc the
times required for a complete decay begin to be comparable to the Hubble time,
in good agreement with the results by M05, and also that it is quite possible
that M54 have reached the very center of Sgr only in recent times, if it was
born sufficiently far away from the center. 
The analytical computations by M05
also indicated that clusters with a mass similar to the other GC residing in the
main body of Sgr (Ter~7, Arp~2, Ter~8, $M\le 2.\times 10^4 M_{\sun}$), would
have infinitely long decay times (i.e. stable orbits) 
if they born at $r\ga 1$ kpc from
the center of Sgr. As the closest of these clusters lie at $\sim 2.9$ kpc from
the center of Sgr M05a concluded that all of them are on stable orbits.
Simulation bh02\_a is intended to verify this conclusion, by letting evolve a
BH as massive as the most massive Sgr cluster (except M54, i.e. Arp~2, 
$M\la 2.\times 10^4 M_{\sun}$, adopting the same $(M/L)_V$ of M54, wich seems
appropriate given the age and metallicity of the cluster) from a starting
point $X_0=2.0$ kpc and a low initial velocity ($V_{Y,0}=4.0$ km/s). After 6
Gyr no sign of decaying is noticed, the orbit is absolutely stable, in
excellent agreement with the results by M05a.

\item The effect of dynamical friction is to progressively decrease the 
radius, the period and the eccentricity of the orbit. 
The mean velocity, on the other hand,  appears to drop suddenly in the last
phases of the decay, possibly when the point-mass particle reaches the dense 
central cusp
of the host NFW halo. This may suggest that the presence of a central cusp may
be a crucial ingredient to lead to the observed phase-space coincidence between
M54 and Sgr,N, in particular if the recent results by \citet{sanchez} and
\citet{read} are considered. This suggestion is worth being followed up with
specific high-resolution simulations adopting different kinds of host galaxy
halo models (such as King models, for instance).
Recent studies indicates that dynamical friction is probably much 
less efficient in cored than in cusped structures
\cite[see][]{sanchez,read,goerdt}.

\end{enumerate}

\begin{figure}
\includegraphics[angle=0,width=\hsize]{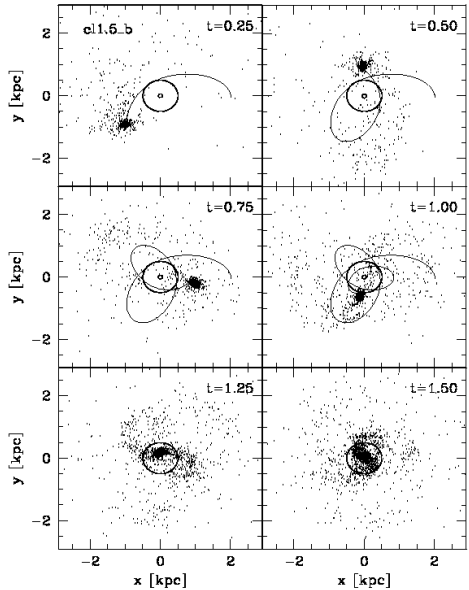}
\caption{Evolution of the M54 cluster model within the NFW model of
the Sgr galaxy, for the simulation cl1.2\_N (see Tab.~\ref{Tnbody}).
In each snapshot of the x,y distribution we plot only the points of the M54
model, for clarity. Concentric circles of radius 60 pc and 500 pc  are plotted
in each panel, for reference. In the top 4 panels we also plotted the orbit
of the bh1\_N model, to show the approximate path of the center of mass of the
cluster model.
\label{xy}}
\end{figure}

To check the impact of our adoption of a point-mass model of M54, 
we repeated the simulation bh1\_N and bh15\_b adopting a $10^4$ particle 
King model resembling
the real cluster as much as possible ($W_0=8.0$, $r_t=60$ pc; 
simulations cl1.2\_N and cl15\_b of Tab.~\ref{Tnbody}); 
in the case of cl1.2\_N the mass was adjusted
to $M=1.2\times10^6 ~M_{\sun}$ to have a central l.o.s. velocity dispersion of
$\sigma_0\simeq 15$ km/s. 
All the final results are fully consistent with those
obtained in simulations with a point-mass model for M54, for the purposes of the
present study, hence we do not further
distinguish between these two classes of models. 
To illustrate the typical behavior of these live M54 models in Fig.~\ref{xy}
we show a series of snapshots  of
the evolution of the cluster model of simulation cl15\_b.
It is interesting to note that in this simulation (as well as in the cl1.2\_N
one) the cluster suffers strong tidal disruption from the host halo and the final
relic sitting nearly at rest at the center of the halo is probably unbound, at
odds with the real case. The very fact that the actual cluster is still bound
may provide useful constraints on the mass distribution within the Sgr galaxy,
that can probably be explored with a systematic suite of N-body simulations.

\section{Summary and Discussion}

We present the results of a large radial velocity survey of stars in the nucleus
of the Sgr galaxy and in the globular cluster M54, which lies at the center of
the nucleus itself. From high signal to noise Keck-DEIMOS and VLT-FLAMES 
spectra, we obtained accurate radial velocities ($\epsilon_{V_r} \simeq 2$ km/s) 
and   metallicities from the Ca triplet for 1152   
candidate RGB/RC stars of M54/Sgr,N. Selecting by the position in the CMD,
by radial velocity and by metallicity, we obtained two reliable 
and clean samples
of 321  Sgr,N members and 425  M54 stars and we used these
samples, further cleaned from $3\sigma$ velocity outliers, 
to study the kinematics of the two systems.

In support of this analysis we used archival HST/ACS images to refine the
estimates by M05a of the structural parameters of Sgr,N.

We also performed a suite of N-body simulations of the orbital decay (driven by
dynamical friction) of a massive cluster into a host dwarf spheroidal galaxy,
mimicking the M54 - Sgr system. We focused our attention on the possibility to
realize a phase-space match between the infalling cluster and the density
cusp of the host NFW halo as close as that observed with M54 and Sgr,N.

The main results of our analyses can be summarized as follows:

\begin{enumerate}

\item We have obtained new, much more reliable estimates of the total luminosity
      and of the typical size of Sgr,N, confirming that it is significantly less
      luminous than M54 and that it has a different luminosity distribution with
      respect to the cluster.

\item The systemic radial velocities of M54 and Sgr,N are identical within 
      $\sim \pm 1$ km/s. Coupling this result with those by M05a, 
      it appears that the
      two systems coincide in phase-space to within the observational errors, 
      at least for what concerns position in space and radial velocity.

\item The velocity dispersion profile of M54 is in good agreement with the
      theoretical profile of the King model that best fits its surface
      brightness, at least over a range of 30 core radii ($r\le 3.5\arcmin$), 
      but possibly also beyond this radius (see Sect.~4.2). 
      In particular, the
      velocity dispersion drops from $\sigma=14.2$ km/s at $r=0$, to 
      $\sigma\simeq 5.3$ km/s at $r\sim 3.5\arcmin$. 
      The observed velocity profile strongly suggests that M54 behaves as 
      an ordinary, self-bound globular cluster, at least in its innermost 
      and densest region. In our view, this indicates that even if 
      at the present epoch the motion of its stars is driven by the potential
      produced by the overall mass distribution within Sgr,N (see below),
      its velocity profile keeps memory of its original nature, that is
      an ordinary massive globular cluster orbiting within the Sgr galaxy. 
      A turn-over of the profile to 
      $\sigma=8.8$ km/s around the tidal radius of the cluster may suggest that
      some degree of tidal disruption is occurring in the outskirts of the
      cluster; however the alternative hypothesis that the sample is 
      contaminated by metal-poor Sgr stars
      at large radius is also viable and seems more
      likely, at the present stage (Sect.~4.2).
      
\item The velocity dispersion profile of Sgr,N is consistent with being flat at
      $\sigma\simeq 10 $ km/s over the whole considered radial range ($r\le
      9\arcmin$). The profile is hardly compatible with any realistic King
      model roughly reproducing the observed surface brightness profile of
      Sgr,N. More generally, {\em the fact that the velocity profile is flat in a radial range in
      which the surface density declines by a factor of $>1000$ is strongly
      uncompatible with any mass-follows-light model} \citep{gil}.
      Conversely, a realistic NFW model reasonably reproduces the
      observed velocity profile and, in principle, does not conflict with the
      presence of an overdensity of baryons at its center, like Sgr,N.
      However, it has to be recalled that the NFW model does not make any
      definite prediction on the SB profile of the embedded stellar nucleus and
      the compatibility of the observed SB profile of Sgr,N with the NFW model 
      that that fits the velocity profile is not established. 
      
\item M54 and Sgr,N have definitely different kinematical properties. In
      particular, the velocity dispersion profiles are very different: in the
      radial range $1.5\arcmin < r < 6.5\arcmin$ the statistical significance of the
      difference in the velocity distribution is very large.

\item We have provided observational evidence that the velocity dispersion 
       profile of Sgr remains flat from $r\simeq
      1\arcmin$ to $r\simeq 100\arcmin$ and that there is no apparent transition
      in the velocity profile corresponding to the onset of the stellar nucleus,
      at $r\la 10\arcmin$.
      This fact as well as those listed above strongly suggest that Sgr,N and
      M54 had independent origins, as we would expect that, if the cluster 
      provided the mass seed to
      collect the Sgr gas that later formed the metal-rich nucleus, Sgr,N stars
      would have shown a declining velocity dispersion profile, compatible with
      a mass-follows-light distribution.
      
\item Our N-body simulations that follow the orbits of massive clusters 
      (1-2 $\times 10^6 M_{\sun}$, representing M54) within different NFW halos 
      (representing the Sgr galaxy) show that for a large range of initial 
      distances and relative velocities, the orbit of the cluster 
      decays completely by dynamical friction within 3 Gyr, at most. Moreover, at
      the end of the simulations, the cluster is perfectly concentric with the
      cusp of the host halo (within the resolution of the simulation) and the
      difference in average velocity is always less than $\sim 2$ km/s. Hence,
      the observed phase-space coincidence between M54 and Sgr,N can be
      naturally explained by the ``dynamical friction hypothesis'' (M05a).
      
\item According to FC06 the mass of Central Massive Objects, independently if
      they are Black Holes or stellar nuclei, is  $\simeq \frac{3}{1000}$ of
      the mass of the host galaxy ($\frac{M_{CMO}}{M_{gal}}=0.003$). Using our
      estimates for  the total mass of Sgr,N and M54 we obtain the following
      estimates for the total mass of the Sgr galaxy, depending on 
      whether we assume is the
      M54, Sgr,N or the sum of the two  (as we would do if we observed the
      system at the distance of Virgo cluster) is the Central Massive Object:
      $M_{Sgr}=1.0\times10^9~M_{\sun}$, $M_{Sgr}=2.1\times10^9 ~M_{\sun}$, and 
      $M_{Sgr}=3.0\times10^9~M_{\sun}$, respectively. 
      These values are in very good agreement
      with  the independent estimates by \citet{maj} that ranges from 
      $M_{Sgr}=5.8\times10^8 ~M_{\sun}$ to $M_{Sgr}=6.9\times10^9~M_{\sun}$.
      Hence both Sgr,N and M54 singularly or taken together 
      have a mass compatible with being the CMO of the Sgr 
      galaxy. 

\item Both Sgr,N and M54, as well as the combination of the two, when placed in
      the $M_V$ vs. log$r_h$ diagram lie in a region that is populated by
      globular clusters {\em and} galactic stellar nuclei.  
      They are also compatible with the Color-Magnitude
      relation of nuclei shown by FC06 and with the $M_V$ vs. $\sigma$ 
      relation satisfied by globular clusters, nuclei and UCDs 
      \cite[see][]{geha,evsti}, as shown in Fig.~\ref{mvsigma}
      Hence the structure and dynamics of Sgr,N, M54 and their combinations
      are fully compatible with other galactic nuclei.
           
\item A detailed study of the mass profile of Sgr,N and M54 using the
      Schwarzschild method \cite[see][and references therein]{rix,ven} is
      currently ongoing and the results will be presented elsewhere (Ibata et
      al., in preparation). 	   
	   
\end{enumerate}

These findings lend very strong support to the scenario proposed
by M05a to explain the M54/Sgr,N system: the nucleus of the galaxy formed
{\em in situ}, at the bottom of the potential well of the Sagittarius
galaxy; the {\em globular cluster} M54 was independently driven to 
the same site by dynamical friction. 
As a complement of the above conclusions, it must be recalled that
the stellar population that dominates Sgr,N formed several Gyrs later than
M54 \citep[see][and references therein]{paul,sl97,ls00,mic_con,siegel}. 
In the
present context this does not appear particularly relevant, but it should be
kept in mind that M54 can have reached the very center of Sgr {\em after}
Sgr,N formed its stars 
\cite[within the last $\sim 5-9$ Gyr][]{mic_con}, or even {\em before}
this, depending on the birthplace of the cluster within the Sgr
galaxy. 
In any case it is very likely that both the cluster and the 
processed Sgr gas
were {\em independently} driven - by different mechanisms - 
to the bottom of the potential 
well of the Sgr galaxy, i.e. to the center of its Dark Matter halo.
A conclusive proof that M54 was driven to its present position by dynamical
friction could be provided by the successful detection of genuine extra-tidal
stars at large distances from the cluster center as envisaged in Sect.~4.3,
while, as said their non-detection will not disprove the above scenario.

As a possible alternative to this view (or as an extreme version of it) it can
be conceived that M54 formed at the bottom of the overall potential well
of the Sgr galaxy since the beginning. In this framework M54 and Sgr,N are just
the results of two subsequent episodes of star formation both occurring at the
very center of Sgr, the second from  enriched gas that was infalling on very
different orbits with respect to the gas that formed M54, thus resulting in
different final stellar kinematics. We regard this possibility as much
less likely with respect to the ``dynamical-friction'' scenario depicted above. 
First, if M54 is considered as part of the (metal-poor) field population of Sgr,
its presence
would be at odd with the global metallicity gradient observed by \citet{alard},
\citet{sdgs2}, \citet{sl97} and others in Sgr, as well in all dwarf
spheroidal galaxy studied to date 
\cite[more metal rich and younger populations are preferentially found at 
the center of dSph's][]{harbeck}. 
While dE nuclei have been indicated as a
possible exceptions to this trend, the ingestion of large
metal poor globular clusters in pre-existing metal rich nuclei seems one 
possible natural way to reconcile the generally observed metallicity
gradients and the presence of nuclei that are bluer than their parent galaxy
\citep{lotz4}.  
Second, as both M54 and Sgr,N would have {\em formed} from gas falling into 
the same potential well, the reason for the different kinematics remains to be 
explaned, while it is a natural outcome if M54 formed elsewhere 
as a classical globular cluster. 

\begin{figure}
\plotone{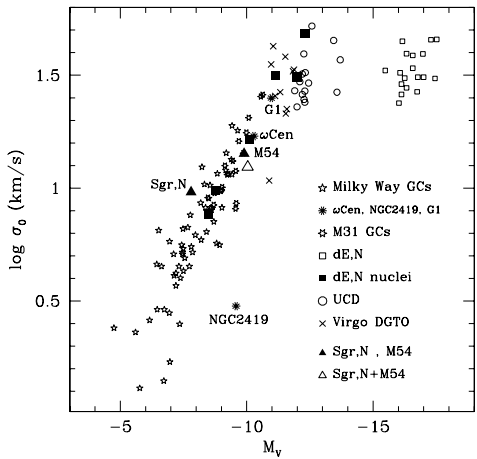}
\caption{Sgr,N and M54 (filled triangles) in the $M_V$ vs. $\sigma$ plane
\cite[see][]{evsti}. Both systems lie on the locus common to 
globular clusters, dE nuclei, Dwarf-Globular Transition Objects 
\cite[DGTO][]{hase}, and, possibly, UCDs. The same is true for their
combination (Sgr,N+M54, open triangle), 
i.e. for a system having luminosity equal to the sum of the 
luminosities of M54 and Sgr,N and having $\sigma$ as estimated from an 
integrated spectrum obtained with a $1\arcsec$ slit from the distance of 
the Virgo cluster of galaxies ($\sigma\simeq 12.4$ km/s). 
Data for Galactic globulars
are from \citet[][$M_V$]{macsyd} and \citet[][$\sigma$]{tad}; for M31 globulars
we took integrated magnitudes from the Revised Bologna Catalog \citep{rbc} and
$\sigma$ from \citet{djor}; data for dE and dE,N are from \citet{geha,geha2}; 
UCDs data are from \citet{ucd} and \citet{evsti}; data for DGTO are from
\citet{hase}. Some remarkable bright clusters have been labeled, for reference
\cite[see][]{lucky}.
\label{mvsigma}}
\end{figure}

\subsection{The process of galaxy nucleation}

If we take for demonstrated the above concluding remarks, we can ask
what we have learned about the process of galaxy nucleation from the case
studied here. Concerning the two main mechanisms that have been 
put forward in the literature, i.e.
(a) formation of the nucleus by infall of globular cluster(s) to the center of
the galaxy, or (b) {\em in situ} formation by accumulation of gas at the center
of the potential well and its subsequent conversion into a stellar overdensity
\citep[see Sect. 1. and][]{grant,cote6}, the main conclusion that can be drawn
from the case of Sgr is that both channels are viable and actually both have
been at work ``simultaneously'' in Sgr. 

The present analysis has shown that a stellar nucleus made of the typical
material that dominates the baryonic mass budget of Sgr is present in this
galaxy, independent of M54, as it display the same flat velocity dispersion
profile as the whole core of Sgr, much different from that of the cluster. 

In a likely scenario, the enriched gas from previous generations of 
Sgr stars accumulated at the bottom of the overall potential well of the galaxy, 
until star formation transformed it into a stellar nucleus whose surface 
brightness is $\ga 100$ times larger than in the surrounding Sgr core, a 
substructure within a larger galaxy. 
On the other hand, M54 is a (relatively) ordinary massive, old
and metal poor globular cluster. Independent of its birthplace within the
early Sgr galaxy, its mass and the density of the surrounding medium of the host
galaxy drove it to the bottom of the Sgr potential well, by dynamical friction.
During its trip to the densest central region of Sgr, the dense cluster 
managed to survive the tidal force of the host galaxy, hence it reached the
present position  as a (partially?) self-bound stellar system.

While the mass budget at the center of Sgr is probably dominated by the 
underlying DM halo, M54 dominates the overall light distribution : 
an observer taking photometry and/or spectra of the unresolved 
nucleus of Sgr from a distant galaxy (say, a galaxy in the Virgo cluster) would
find that the object looks like a bright and blue globular cluster; 
the integrated velocity
dispersion would not show any peculiar feature revealing the composite nature of
the observed nucleus (see Fig.~\ref{mvsigma}, above). 
Probably, it would be impossible to disentangle the two systems from 
the integrated light\footnote{It would be interesting to check if there is 
some spectral feature that may reveal the composite nature of the ``system'', in
the present case. We plan to do this in the future by combining properly 
scaled synthetic spectra representing the light output of M54 and Sgr,N.}.

Finally, the Sgr case seems to support the observed ubiquity of nuclei (see
Sect.~1. and C06). 
The Sgr galaxy was able to
form a sizable nucleus ``twice'' and with two different formation channels:
if either of the two channels had not been viable for some reason, the galaxy
would have ended up with a nucleus in any case. The fact that Sgr is the only
case of galaxy with a clear nucleus among those classified as dwarf spheroidals
in the Local Group may suggest that the progenitor of Sgr was in fact a brighter
dE or disc galaxy that has been transformed into a dSph by the interaction
with the Milky Way \cite[][M05b, and references therein]{maj,mayer}

\subsection{Suggestions for further investigations}

The results presented in this paper suggest several interesting
lines of research that we did not follow up for practical reasons.
However we feel that it is worth briefly mentioning some of them here, as a
possible starting point for future studies.

\begin{itemize}

\item The results presented by
\citet{read}, \citet{goerdt} and \citet{sanchez}, suggest that the complete 
decay of a massive
cluster to the very center of the host galaxy is much easier and faster if a
central density cusp is present. It is even possible that a central cusp is
actually {\em required} to bring a cluster down to the very bottom of the
overall potential well. If this is the case, the position of M54 within Sgr,N
would support the existence of a central cusp in actual DM halos, a point that
has been questioned by several authors \cite[see][and references
therein]{sanchez}. 
The ``complete infall'' of a massive cluster may need a NFW cusp and,
simultaneously, it may transform the cusp into a core by transferring orbital
energy and momentum to the surrounding ``medium'',
thus possibly providing a self-regulating mechanism that simultaneously
prevent the further decay of other clusters \citep[that, in general cases, 
would be quite difficult, given the expected 
decay times, see][and references therein]{herna,oh,read,sanchez,goerdt}.
In this context, it may be worth to recall also the work by
\citet{stri}, whose results militates against the presence of a large-size
core in the Fornax dSph, and by \citet{boy} on the resilience of cuspy halos
in major mergers. 
These ideas seems worthy of detailed theoretical follow up.

\item It has been suggested several times that bright and metal poor globular clusters
may be of cosmological origin \citep[see][and references therein]{BS,sills}. 
If this
is the case they may be very intimately linked to the earliest phases of
formation of galaxies and there may have been plenty of opportunities for most
of them to become the nuclei of some dwarf galaxy. The tidal stress that they
suffered during their infall to the center of their host galaxies may be at the
origin of the larger half-light sizes that are observed in those of them that
have been suggested as possible nuclear remnants of ancient dwarf galaxies
\citep{frebland,lucky,macsyd}. This kind of scenario may be explored
with dedicated N-body simulations, possibly including gas and stars.
It would be interesting also to consider in detail the results presented here
in relation with the scenario for the origin of globular clusters recently 
discussed by \citet{bok07}.

\item All the analysis presented in this paper have been performed within the
standard Newtonian gravitation theory and dynamics. 
It may be worth to consider the
observational scenario emerged from this study also in the framework of Modified
Newtonian Dynamics paradigms 
\citep[MOND, see][and references therein]{mil,sand}, even if it may not
necessarily be an ideal case.
The transition between ordinary Newtonian regime and MOND regime occurs around
$r\simeq 4.5\arcmin - 6.5\arcmin$, depending on the actual stellar mass of 
M54+Sgr,N, that is in a range covered by our data.

\item There is little doubt that the final fate of the Sgr galaxy will be 
its
complete tidal disruption. Once the large scale stellar body of Sgr 
will be completely
dispersed into the Galactic halo, the final remnant of this (once) relatively 
large galaxy would be a faint nucleus embedding a bright globular cluster.
An observer lacking any knowledge of the origin of this object would conclude
that it is a very bright and peculiar globular cluster, dominated by a metal
poor population ($[Fe/H]\sim -1.5$, possibly with some spread) 
but also including a small fraction 
($\sim$ 10\%) of metal rich stars (with average $[Fe/H]\sim -0.4$). Also the
abundance patterns would appear different: for instance, the metal poor (M54)
stars would appear as moderately $\alpha$-enhanced \citep{brown}, while metal
rich stars (Sgr,N) would have solar or sub-solar $[\alpha/Fe]$ ratios
\citep{luves}.
The radial
velocities would reveal that the metal rich and metal poor stars have {\em 
the same systemic velocity}, but {\em different velocity dispersion profiles}
and {slightly different density profiles}, possibly with some rotation in the
metal poor component. However
the dominance of metal-poor (M54) stars would be so high that there would be no
hint of the presence of a Dark Matter component. The half-light radius of
the system would appear slightly larger with respect to ordinary globulars
\citep{macsyd,lucky}. 
Most of these likely characteristics
of the future remnant of Sgr seem to have a counterpart in the
widely studied and mysterious stellar system $\omega$ Centauri 
\citep[see][and references therein]{norris,omega,ele1,ele2,ele3,ven}, 
that has been proposed as the
possible remnant of a nucleated dwarf elliptical since \citet{freeman}.
While there are also noticeable differences between the two cases, the analogy
seems very intriguing and potentially 
powerful\footnote{It is interesting to recall that $\omega$ Cen shares with 
Sgr some remarkable chemical peculiarities. In particular, 
stars of comparable
metallicity in the two systems are strongly enhanced in s-process 
elements and the ratio of heavy s-process to light s-process elements [hs/ls] 
is very similar. Furthermore, $\omega$ Cen and Sgr are, up to now, the
only stellar systems known to have deficient [Cu/Fe] ratios with respect to 
the trend in the Galactic disk and halo \citep{mcw}}. 
It is possible that at least
some of the observational features of $\omega$ Centauri that appear so difficult
to explain may find their natural place within a scenario like the one
described above.

\end{itemize}

\acknowledgments

M.B. acknowledges the financial support of INAF through the Grant  PRIN05 CRA
1.06.08.02. M.B. is grateful to the Observatoire de Strasbourg for the kind
hospitality during the period in which most of the N-body simulations presented
here have been performed.  This work is partially based on observations made
with the European Southern Observatory telescopes (WFI@2.2m) as part of the
observing program 65.L-0463. Partially based on observations made with the ESO
Very Large Telescope (VLT/FLAMES) as part of the observing program 075.D-0075.
Partially based on Advanced Camera for Surveys (ACS) observations collected
with the Hubble Space Telescope within the programme GO 10755. This research
has made use of the NASA Astrophysics Data System Abstract Service.


{\it Facilities:} \facility{Keck (DEIMOS)}, \facility{ESO-VLT (FLAMES)},
\facility{MPG-ESO (WFI@2.2m)}, \facility{HST (ACS)}.





\end{document}